\documentclass[aps,twocolumn,prb,superscript,floatfix,superscriptaddress,showpacs]{revtex4-1}
\usepackage{amssymb}

\usepackage{epsf}
\usepackage{epsfig}
\usepackage{graphicx}
\usepackage{dcolumn}
\usepackage{braket}
\usepackage{bm}
\usepackage{amsfonts}
\usepackage{amsmath}
\usepackage{amssymb}
\usepackage{color,soul}
\usepackage{wasysym}
\usepackage{mathrsfs}
\usepackage{natbib}
\usepackage{multirow}
\usepackage[colorlinks=true,%
            linkcolor=blue,%
            urlcolor=blue,%
            citecolor=blue,%
            filecolor=blue,%
            bookmarksopen=true,%
            pdfauthor={J. P. Ramos-Andrade},%
            pdftitle={AB_QD_MBS},%
            pdfsubject={AB_QD_MBS_Manuscript},%
            pdfpagemode=UseOutlines]{hyperref}

\newcommand{\ve}{\varepsilon}

\newcommand{\bea}{\begin{eqnarray}}
\newcommand{\eea}{\end{eqnarray}}

\newcommand{\la}{\langle}
\newcommand{\ra}{\rangle}

\begin{document}

\title{Detecting coupling of Majorana bound states with an Aharonov-Bohm interferometer}
\author{J. P. Ramos-Andrade}
\email{juan.ramosa@usm.cl}	
\affiliation{Departamento de F\'isica, Universidad T\'ecnica Federico Santa Mar\'ia, Casilla 110 V, Valpara\'iso, Chile}
\affiliation{Department of Physics and Astronomy, and Nanoscale and Quantum Phenomena Institute, Ohio University, Athens, Ohio 45701-–2979, USA}
\author{P. A. Orellana}
\affiliation{Departamento de F\'isica, Universidad T\'ecnica Federico Santa Mar\'ia, Casilla 110 V, Valpara\'iso, Chile}
\author{S. E. Ulloa}
\affiliation{Department of Physics and Astronomy, and Nanoscale and Quantum Phenomena Institute, Ohio University, Athens, Ohio 45701-–2979, USA}

\begin{abstract}
We study the transport properties of an interferometer composed by a quantum dot (QD) coupled with two normal leads and two one-dimensional topological superconductor nanowires (TNWs) hosting Majorana bound states (MBS) at their ends. The geometry considered is such that one TNW has both ends connected with the QD, forming an Aharonov-Bohm (AB) interferometer threaded by an external magnetic flux, while the other TNW is placed near the interferometer TNW.  This geometry can alternatively be seen as a long wire contacted across a local defect, with possible coupling between independent-MBS. We use the Green's function formalism to calculate the conductance across normal current leads on the QD. We find that the conductance exhibits a half-quantum value {\em regardless} of the AB phase and location of the dot energy level, whenever the interferometer configuration interacts with the neighboring TNW. These findings suggest that such a geometry could be used for a sensitive detection of MBS interactions across TNWs, exploiting the high sensitivity of conductance to the AB phase in the interferometer.
\end{abstract}

% 73.22.Pr   	Electronic structure of graphene
% 73.20.At 	Surface states, band structure, electron density of states
% 72.80.Vp	Electronic transport in graphene
% 73.63.-b      Electronic transport, nanoscale materials
% 78.67.-n   	Optical properties of low-dimensional, mesoscopic, and nanoscale materials and structures
% 73.43.-f	quantum Hall effects,
% 73.43.Nq	Phase transitions quantum Hall effects
% 05.60.Gg	Transport processes, quantum
\date{\today}
\maketitle
\section{Introduction}

An interesting member of the fermionic family was first proposed by E. Majorana nearly 80 years ago.\cite{Majorana2008} Majorana fermions (MFs) have as a principal feature to be their own anti-particles. Although long-sought after in different contexts, including neutrinos,\cite{Wilczek2009} they seem to have recently materialized in condensed matter systems, appearing as zero-energy excitations in systems with particle-hole symmetry. As MFs are expected to satisfy non-Abelian statistics, they are of interest in quantum computation implementations and are gaining increasing attention.\cite{alicea2016,alicea2011} MFs can be found in several systems, such as the surface of a topological insulator,\cite{Fu2008} and in the vortex core of a $p$-wave superconductor.\cite{Ivanov2001}

Localized MFs, (or Majorana bound states, MBSs) are also notably predicted to be found at each end of a one-dimensional (1D) semiconductor nanowire with spin-orbit interaction (SOI) placed in close proximity to an $s$-wave superconductor and in a magnetic field,\cite{Lutchyn2010,Oreg2010} or at the ends of a chain of magnetic impurities on a superconducting surface.\cite{NadjPerge2014,Ruby2015} These 1D physical systems can be seen as implementations of the Kitaev chain,\cite{Kitaev2001} and are known as topological nanowires (TNWs). Mourik \emph{et al.} \cite{Mourik2012} reported the first observation of Majorana signatures in a semiconductor-superconductor nanowire, built of InSb (indium antimonide) wires and proximityzed by NbTiN (niobium titanium nitride). Several others groups reported also zero-bias conductance peaks in similar hybrid 1D devices.\cite{Deng2012,das2012,Churchill2013,rokhinson2012} MBS pairs present in the same 1D heterostructure are
expected to interact with a coupling strength $\ve_M\propto\exp(-L/\zeta)$, as function of the wire length $L$, where $\zeta$ is the superconducting coherence length.\cite{Kitaev2001} This dependence on wire length was probed in recent experimental work, verifying expectations.\cite{albrecht2016exponential}

Ever since the first observations, the challenges identified in solid state systems involving MBSs have included the detection and manipulation of these states to explore quantum entanglement and eventual quantum computing engineering.\cite{alicea2012} One of the exciting and promising branches in this context is the study of the interplay between MBSs and other nanostructures, such as quantum dots (QDs).\cite{YuXianLi2013,Silva2016,Baranski2017} A clear signature of the presence of an MBS in a system was established by Liu and Baranger as a half-maximum conductance at the Fermi energy (zero-energy point), $\mathcal{G}(\ve=0)=\mathcal{G}_\text{max}/2=e^{2}/2h$, across current leads with an embedded QD.\cite{Liu2011} Subsequently, Vernek \emph{et al.}\cite{Vernek2014} showed that this signature is completely unaffected by changes in QD energy level, as it shifts across the Fermi energy. This behavior was recently measured in a QD coupled to an InAs-Al nanowire heterostructure, exhibiting the MBS pinned state at zero energy.\cite{Deng2016} Interferometer configurations using QDs with side-coupled MBS in their arms have been considered to study interference phenomena, including the Fano
effect.\cite{Ueda2014,Xia2015,EnMing2014,Ricco20162,Zeng2016} Possible MBS implementations for storage information,\cite{PhysRevB.93.165116} and to allow manipulation of MBSs to perform non-Abelian operations have been discussed,\cite{Flensberg2011} including the use of flux-controlled quantum computation.\cite{Hyart2013} Moreover, manipulation of these states can be achieved by external magnetic fields over straight/circular magnetic adatom chains\cite{li2016} or through the construction of protocols in an $X$-shaped junction.\cite{Aristov2016}

In this work, we explore the interaction between two nanowires in the topological phase, hosting MBSs at their ends,
via the conductance across a QD embedded between normal current leads.
The arrangement exploits the high phase sensitivity of the conductance through a system with MBSs.  The system can be seen to represent an independent TNW close to the TNW in the AB interferometer, or perhaps more typically, as a long wire with an intermediate contact to the interferometer.  As such, the phase and conductance monitoring can be used to probe the interaction between the two TNWs in the device.
To address this problem, we consider an effective low energy Hamiltonian in second quantization and solve the system using a Green's function formalism. Our results show that the conductance through the QD can indeed be used to probe the connection between MBSs belonging to different TNWs. The interaction of MBS is manifested in the observation of regular fermionic or Majorana behavior at the zero-energy point when tuning an Aharonov-Bohm phase away from $\pi/2$.

This paper is organized as follows: Section \ref{secmodel} presents the system Hamiltonian and method used to obtain quantities of interest; Section \ref{secresults} shows the results and the corresponding discussion, and finally, the concluding remarks are in Section \ref{secsummary}.

\section{Model}\label{secmodel}

\begin{figure}[tbph]
\centering
\includegraphics[width=0.4\textwidth]{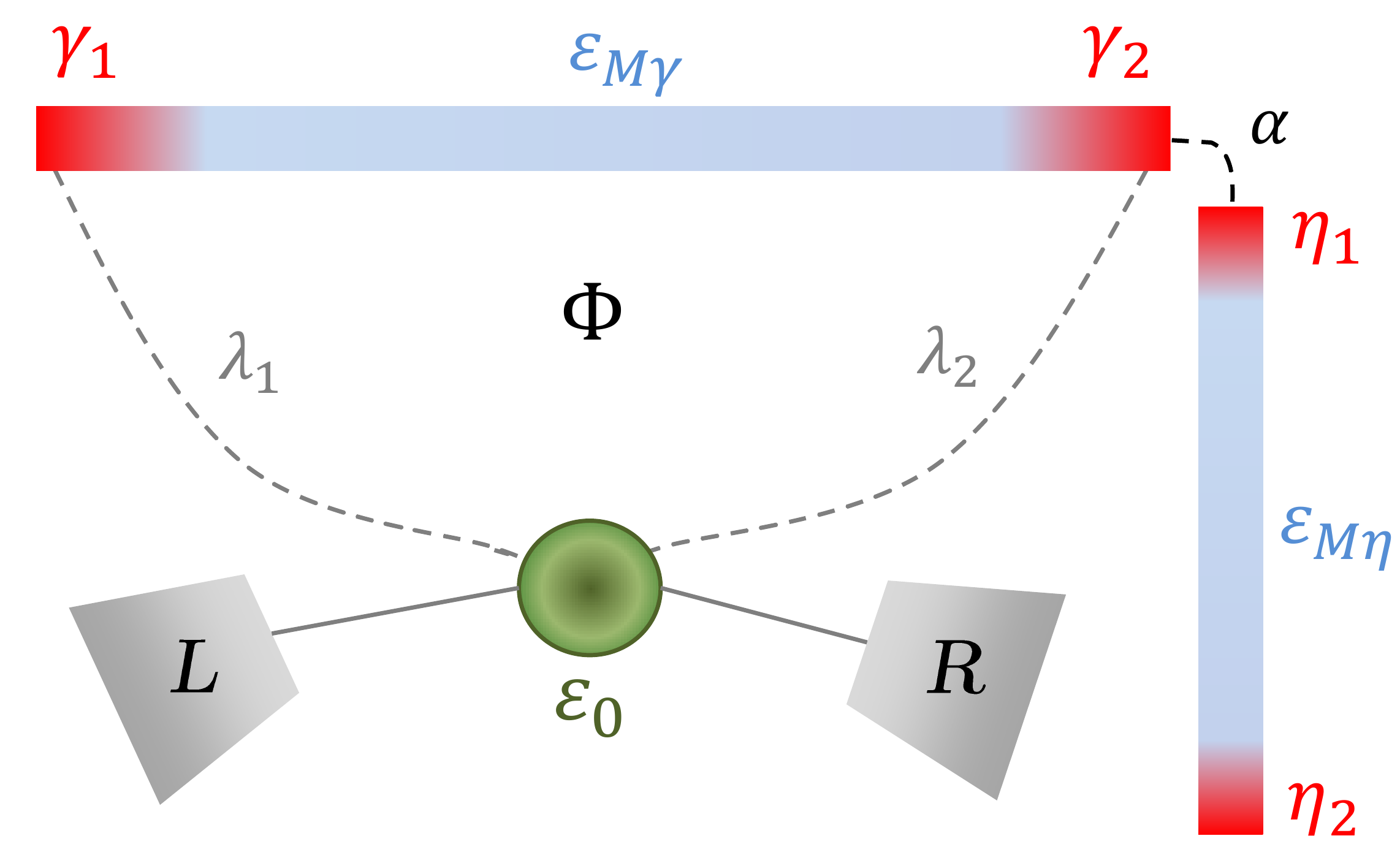}
\caption{Model setup. The system has two topological nanowires (light blue bars) hosting MBSs (red at the end of the bars). A single-level quantum dot (green) is coupled to only one of the TNWs, through the MBS at each end, forming an interferometer pierced by a magnetic flux $\Phi$. One of these MBSs is placed in close  to another MBS, which belongs to the other/neighboring TNW. The QD is also connected to source and drain leads (gray).}
\label{fig1}
\end{figure}

Our system considers two single-mode TNWs hosting MBSs at their ends. One of these TNWs has its ends tunnel-coupled with a single-level QD, forming an Aharonov-Bohm interferometer, and threaded by an external magnetic flux. The other TNW can couple with the interferometer through a direct MBS-MBS coupling across wires. This arrangement could also represent a single TNW where the connection to the interferometer produces a non-topological defect that results  in weakly coupled MBSs across such defect.\cite{Flensberg2010} The QD is also connected to two current leads, labeled left ($L$) and right ($R$), as shown schematically in Fig.\ \ref{fig1}.

We describe the system with an effective low-energy spinless Hamiltonian since only electrons with one spin projection couple to the Majorana states.\cite{ruiz2015} Hence, the Hamiltonian has the form
\begin{equation}
H=H_{\text{leads}}+H_{\text{dot}}+H_{\text{leads-dot}}+H_{\text{MBS}}\,,\label{H0}
\end{equation}
where the first three terms are the regular fermionic (electronic) contributions, given by
\begin{eqnarray}
H_{\text{leads}}&=&\sum_{j,k}\ve_{k}c_{j,k}^{\dag}c_{j,k}\,,\label{Hleads} \\
H_{\text{dot}}&=&\ve_{d}d^{\dag}d\,,\label{Hdot} \\
H_{\text{leads-dot}}&=&\sum_{j,k}t_{j}c_{j,k}^{\dag}d+\text{h.\,c.}\,, \label{Hleadsdot}
\end{eqnarray}
where $c_{j,k}^{\dag}(c_{j,k})$ creates(annihilates) an electron in lead $j=L,R$ with momentum $k$; $d^{\dag}(d)$ does it in the QD, which has a single energy level $\ve_{d}$; and $t_{j}$ is the $k$-independent tunneling coupling between the lead $j$ and the QD.

The last term in Eq.\ (\ref{H0}), $H_{\text{MBS}}$, represents the MBSs and their couplings to other MBSs and the QD. As the Majorana quasiparticles are their own antiparticles, their creation (or annihilation) operator must be real.\cite{Kitaev2001} It follows that they are described by the operators $\xi_{\beta,l}$ that satisfy both $\xi_{\beta,l}^{\dag}=\xi_{\beta,l}$ and the anticonmutator $\{\xi_{\beta,l},\xi_{\beta',l'}\}=\delta_{\beta,\beta'}\delta_{l,l'}$ and $\{\xi_{\beta,l},d\}=0$; here $\beta=\gamma$ for the TNW in the interferometer, and $\beta=\eta$ for the other TNW, and $l=1,2$ (see Fig.\ \ref{fig1}). $H_{\text{MBS}}$ can then written as
\begin{eqnarray}
H_{\text{MBS}}&=&i\ve_{M\gamma}\xi_{\gamma,1}\xi_{\gamma,2}+(\lambda_{1}d-\lambda_{1}^{\ast}d^{\dag})\xi_{\gamma,1}\nonumber \\ &+&i(\lambda_{2}d+\lambda_{2}^{\ast}d^{\dag})\xi_{\gamma,2}+i\ve_{M\eta}\xi_{\eta,1}\xi_{\eta,2} \nonumber \\ &+&i\alpha\xi_{\gamma,2}\xi_{\eta,1}\,,\label{HMBS}
\end{eqnarray}
where $\lambda_{l}$ is the hopping between the QD and MBS $\xi_{\gamma,l}$; $\ve_{M\beta}$ gives the splitting between Majorana modes in the same TNW ($\xi_{\beta,l}$), with $\ve_{M\beta}\propto\exp{[-L_{\beta}/\zeta]}$, where $L_{\beta}$ is the length of the corresponding TNW and $\zeta$ the superconducting coherence length, as mentioned before. The $\alpha$ parameter describes the connection between independent MBSs, \emph{i.e}.\ belonging to different TNWs. In addition, due to the Aharonov-Bohm flux in the system, without loss of generality, we can set $\lambda_{2}=\lambda_{2}^{\ast}$ and $\lambda_{1}\equiv|\lambda_{1}|\exp{(i\phi)}$,\cite{Liu2011} with the phase $\phi=2\pi(\Phi/\Phi_{0})$ given in terms of $\Phi$ the magnetic flux across the system and $\Phi_{0}=h/e$ the flux quantum.

It is helpful to write $\xi_{\beta,1}$ and $\xi_{\beta,2}$ in terms of regular fermionic operators $f_{\beta}^{\dag}$ and $f_{\beta}$, that of course satisfy $\{f_{\beta},f_{\beta'}^{\dag}\}=\{f_{\beta'}^{\dag},f_{\beta}\}=\delta_{\beta,\beta'}$ and $\{f_{\beta},f_{\beta'}\}=0$. One defines \cite{Bolech2007,Leijnse2011}
\begin{subequations}\label{xiefes}
\begin{equation}
\xi_{\beta,1}=\frac{1}{\sqrt{2}}(f_{\beta}+f_{\beta}^{\dag})\,, \label{ximas}
\end{equation}
\begin{equation}
\xi_{\beta,2}=-\frac{i}{\sqrt{2}}(f_{\beta}-f_{\beta}^{\dag})\,. \label{ximenos}
\end{equation}
\end{subequations}
Using Eqs.\ (\ref{xiefes}), the Eq.\ (\ref{HMBS}) transforms to
\begin{align}
H_{\text{MBS}}&=\ve_{M\gamma}\left(f_{\gamma}^{\dag}f_{\gamma}-\frac{1}{2}\right)+\frac{1}{\sqrt{2}}(\lambda_{1}d-\lambda_{1}^{\ast}d^{\dag})(f_{\gamma}+f_{\gamma}^{\dag}) \nonumber \\
&+\frac{1}{\sqrt{2}}(\lambda_{2}d+\lambda_{2}^{\ast}d^{\dag})(f_{\gamma}-f_{\gamma}^{\dag})+\ve_{M\eta}\left(f_{\eta}^{\dag}f_{\eta}-\frac{1}{2}\right) \nonumber \\
&+\frac{\alpha}{2}(f_{\gamma}-f_{\gamma}^{\dag})(f_{\eta}+f_{\eta}^{\dag})\,. \label{HMBSconf}
\end{align}

In order to calculate the transport quantities we used a Green's function (GF) formalism. The retarded GF for two operators $A$ and $B$ in the time domain is defined by
\begin{equation}
\la\la A,B\ra\ra(\tau,\tau^{\prime})=-i\theta(\tau-\tau^{\prime})\la\{A(\tau),B(\tau^{\prime})\}\ra\,, \label{ABtime}
\end{equation}
where $\theta(\tau-\tau^{\prime})$ is the Heaviside function and $\langle...\rangle$ denotes the average over the ground state or thermal average at finite temperature. To obtain the corresponding GF we require, we solve the equation of motion of Eq.\ (\ref{ABtime}) with the full Hamiltonian $H$ and then its Fourier transform to take the equation into the energy domain, so that
\begin{equation}
(\ve+i\delta)\la\la A,B \ra\ra_{\ve}=\la\{A,B\}\ra+\la\la[A,H],B\ra\ra_{\ve}\,, \label{ABenergy}
\end{equation}
with $\delta\rightarrow0^+$, and $[...,...]$ denotes the commutator.

We use the distributive property of the GFs and write the MBS GFs in terms of the fermionic operators $f_{\xi}$ and $f_{\xi}^{\dag}$, such as
\begin{eqnarray}
2\la\la\xi_{\beta,1(2)},\xi_{\beta,1(2)}\ra\ra_{\ve}&=&\la\la f_{\beta},f_{\beta}^{\dag}\ra\ra_{\ve}+\la\la f_{\beta}^{\dag},f_{\beta}\ra\ra_{\ve} \nonumber \\&+&(-)\left[\la\la f_{\beta},f_{\beta}\ra\ra_{\ve}+\la\la f_{\beta}^{\dag},f_{\beta}^{\dag}\ra\ra_{\ve}\right]\,.\nonumber \\ \label{GRMBS}
\end{eqnarray}

With the Green's function at hand, we obtain the local density of states (LDOS) in the QD and the conductance across the leads using the expressions
\begin{equation}
\text{LDOS}_{d}=-\frac{1}{\pi}\text{Im}\la\la d,d^{\dag}\ra\ra_{\ve}\,, \label{LDOSdot}
\end{equation}
\begin{equation}
\mathcal{G}(\ve)=\frac{e^{2}}{h}\Gamma\int\left(-\frac{\partial f_{\text{F}}(\varepsilon)}{\partial\ve}\right) [-\text{Im}\la\la d,d^{\dag}\ra\ra_{\ve}] \, \text{d}\ve\,, \label{condu}
\end{equation}
where $f_{\text{F}}(\ve)$ is the Fermi distribution function, and $\Gamma=\Gamma_{L}+\Gamma_{R}$ is the energy-independent coupling strength between
the QD and the leads in the symmetric case, where $t_{L}=t_{R}$, and $\Gamma_{j}=\pi|t_{j}|^{2}\rho_0$, where $\rho_0$ is the density of states in the leads.

\section{Results}\label{secresults}

\begin{figure}[tbph]
\centering
\includegraphics[width=0.495\textwidth]{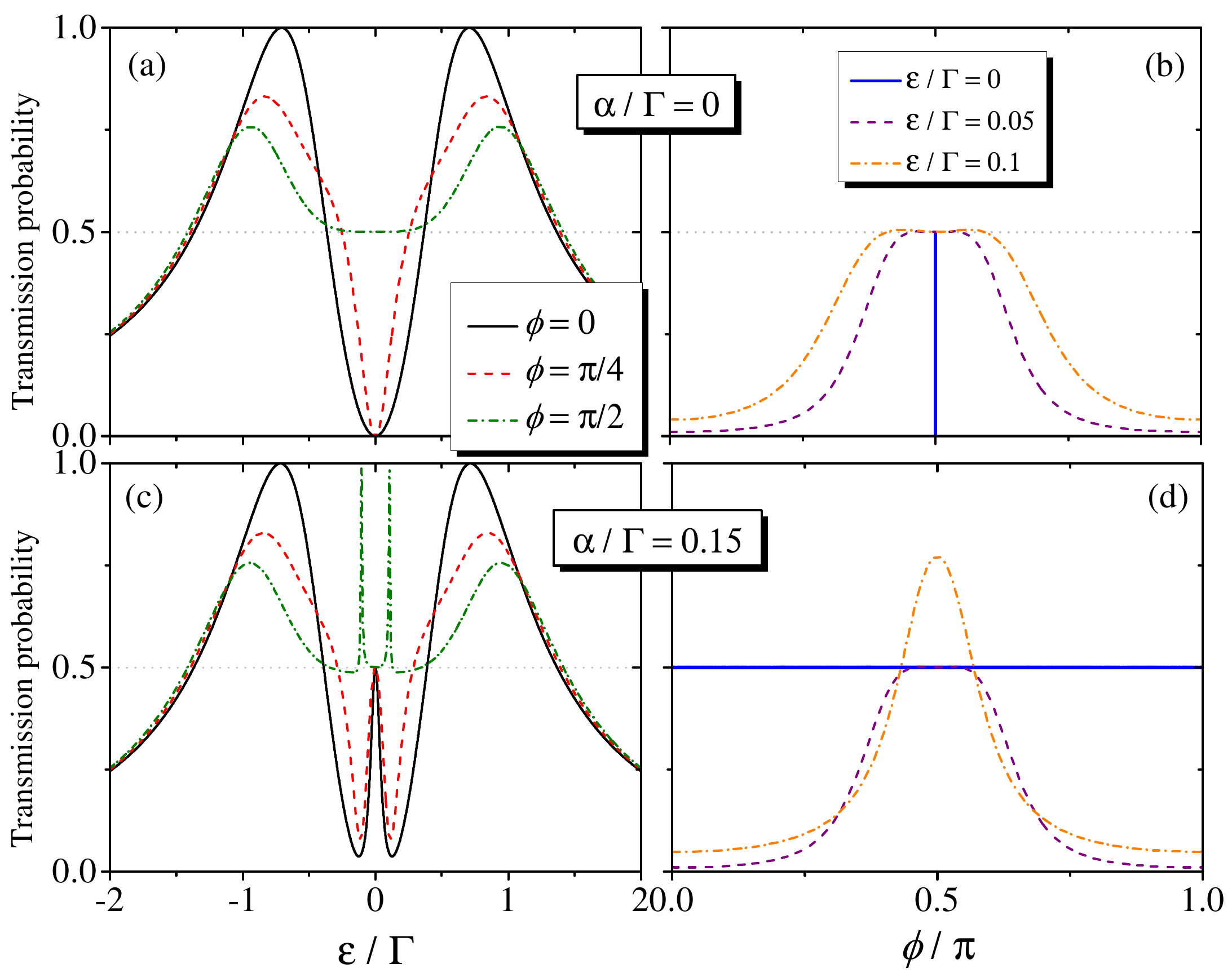}
\caption{Transmission probability through the QD, $\mathcal{T}(\ve)$, as function of energy [(a) and (c)] for fixed Aharonov-Bohm phase values $\phi$, and as function of phase at zero-energy (Fermi level), $\mathcal{T}(\ve=0)$  [(b) and (d)]. $\alpha=0$ is for panels (a) and (b) where the TNWs are not coupled. $\alpha=0.15\,\Gamma$ is for panels (c) and (d). Dotted gray line in (a), (b) and (c) represents the half-maximum conductance. The QD level is at the Fermi level, $\ve_{d}=0$, and $\xi_{M\gamma}=\xi_{M\eta}$=0 in all panels.}
\label{fig2}
\end{figure}

\begin{figure}[tbph]
\centering
\includegraphics[width=0.225\textwidth]{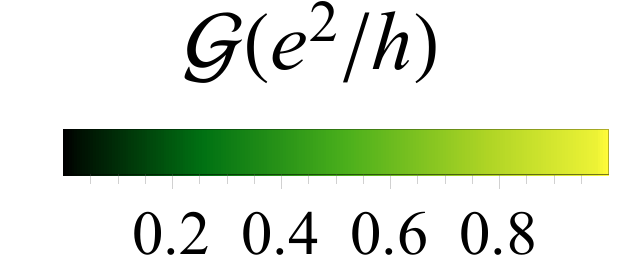}
\includegraphics[width=0.45\textwidth]{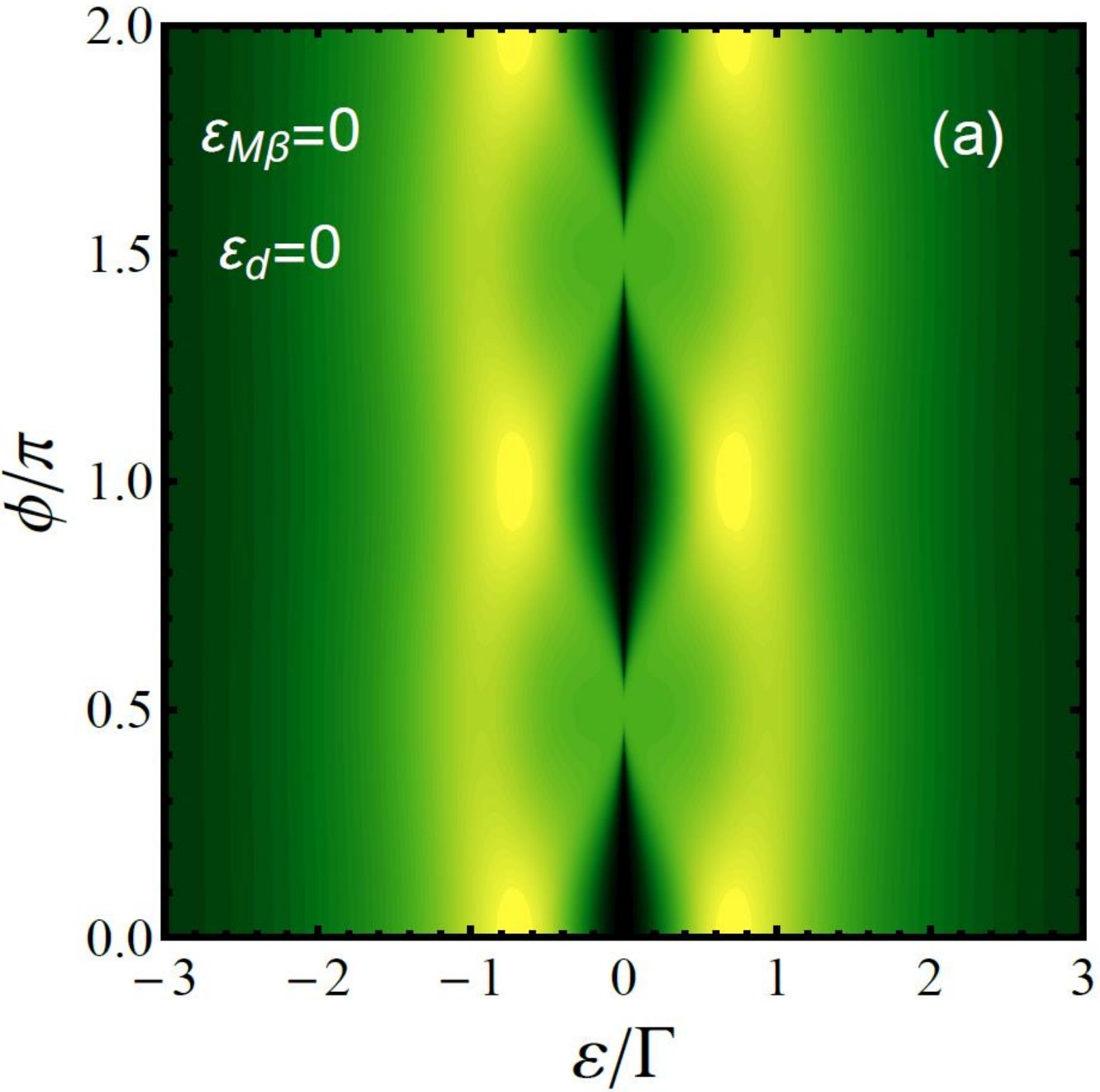}\\
\includegraphics[width=0.45\textwidth]{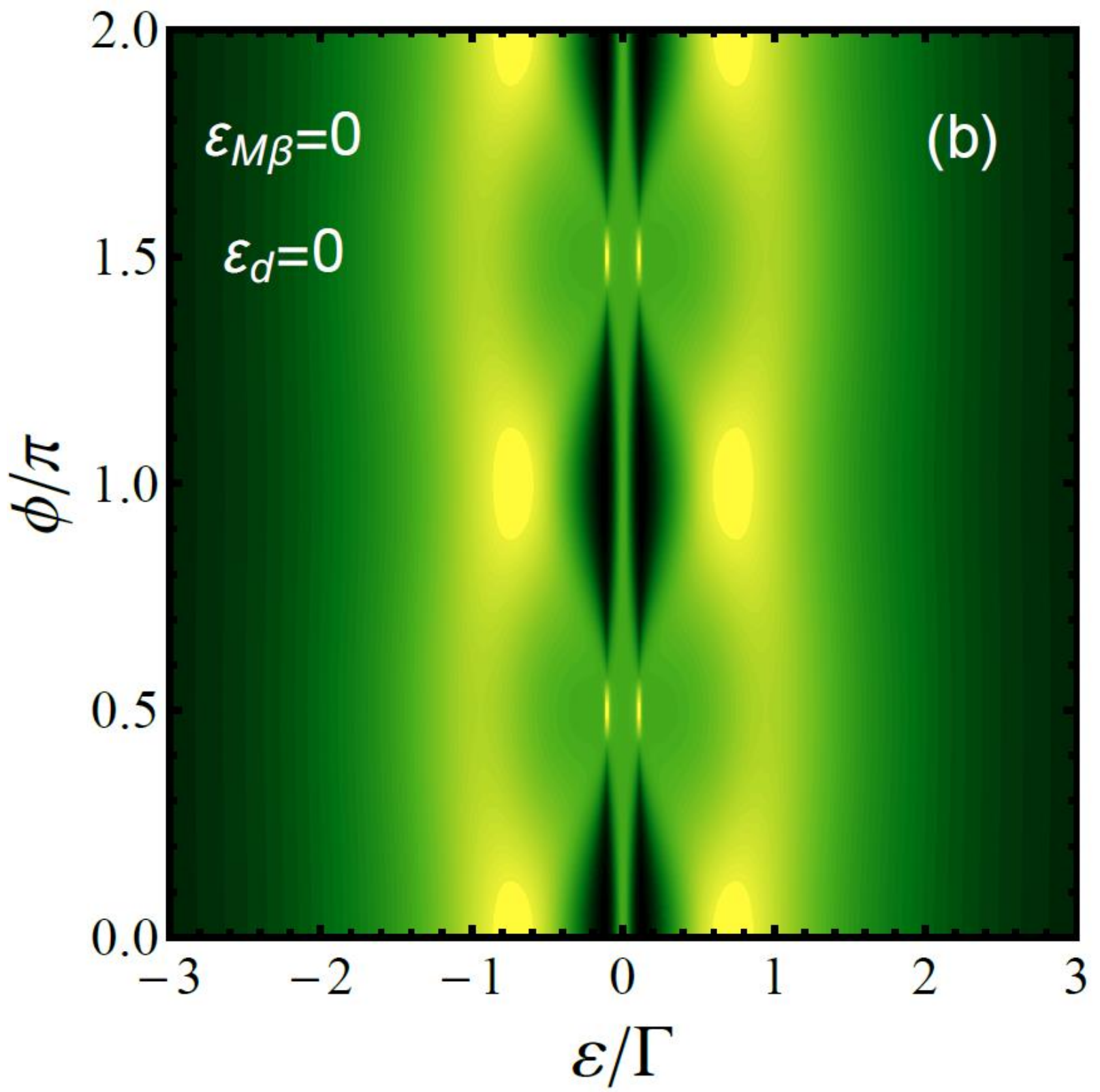}
\caption{Transmission probability map as function of energy and phase.  At $\ve=0$ this is proportional to the zero-temperature conductance. In (a) $\alpha=0$ and (b) $\alpha=0.15\,\Gamma$. The QD energy level is at $\ve_{d}=0$.}
\label{fig3}
\end{figure}

In what follows, we adopt $\Gamma$ as the energy unit. Typical experimental values for $\Gamma$ are few meV.\@ We further fix coupling between the MBS $\xi_{\gamma,l}$ and the QD as $|\lambda_{l}|=\Gamma/2$.  Since typically $T \ll \Gamma$ in experiments, we use the $T=0$ limit of Eq.\ (\ref{condu}), which reduces to
\begin{equation}
\mathcal{G}(\ve_{\text{F}})=\frac{e^{2}}{h}\mathcal{T}(\ve=\ve_{\text{F}})\,,
\end{equation}
where $\mathcal{T}(\ve)=-\Gamma\,\text{Im}\la\la d,d^{\dag}\ra\ra_{\ve}$ is the transmission probability and $\ve_{\text{F}}$ the Fermi energy.
The appendix shows analytical expressions for the appropriate GFs, as well as the conductance, in terms of the system parameters.  We discuss there some of the general features of the conductance, such as the role of the AB flux on different behaviors.  For simpler more intuitive visualization, we show  results for specific cases in what follows.

\begin{figure*}[tbph]
\centering
\includegraphics[width=0.15\textwidth]{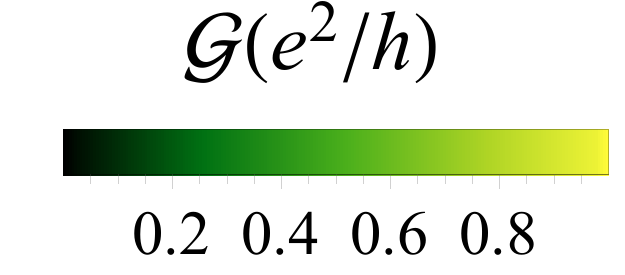}\\
\includegraphics[width=0.33\textwidth]{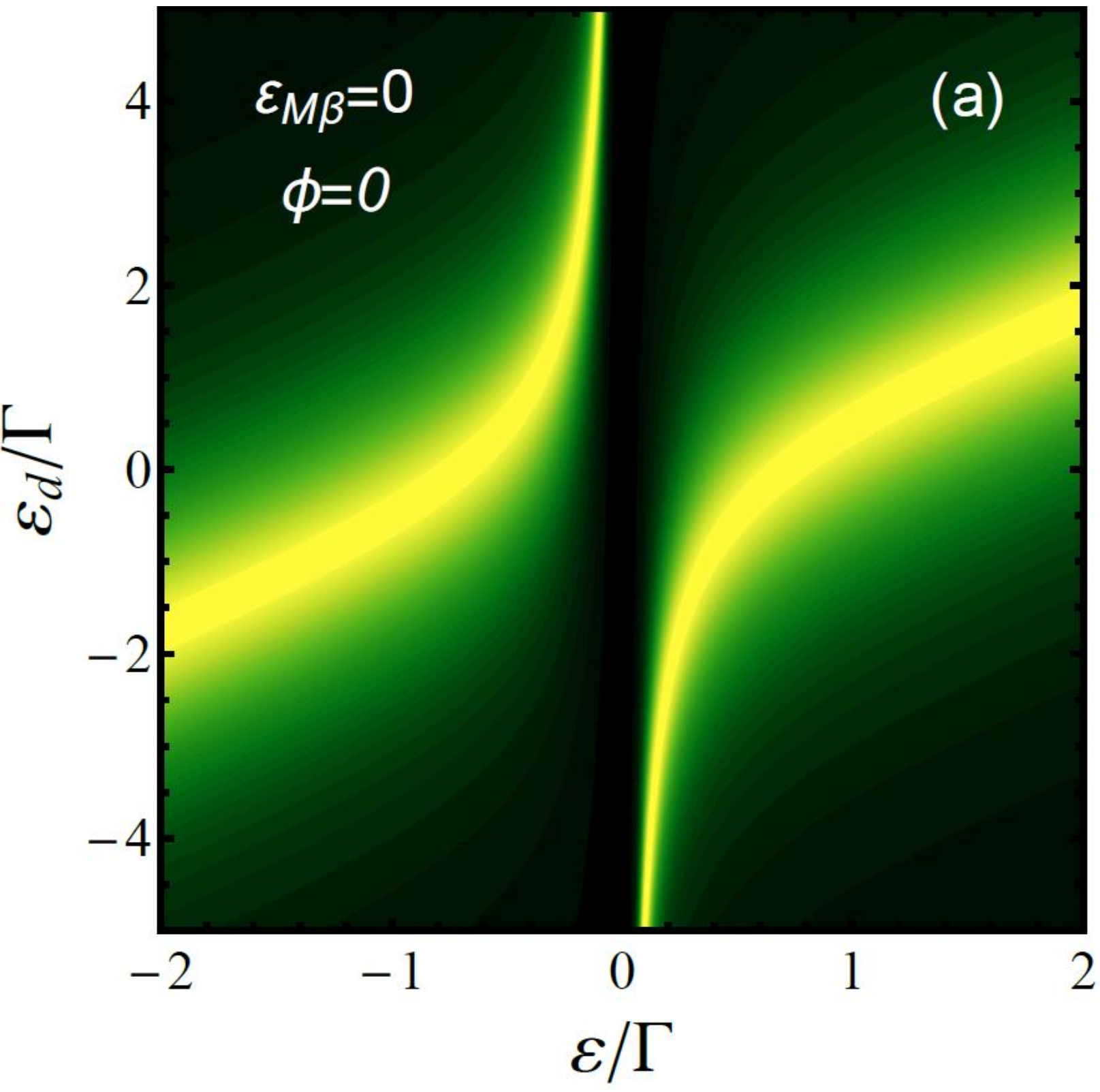}\includegraphics[width=0.33\textwidth]{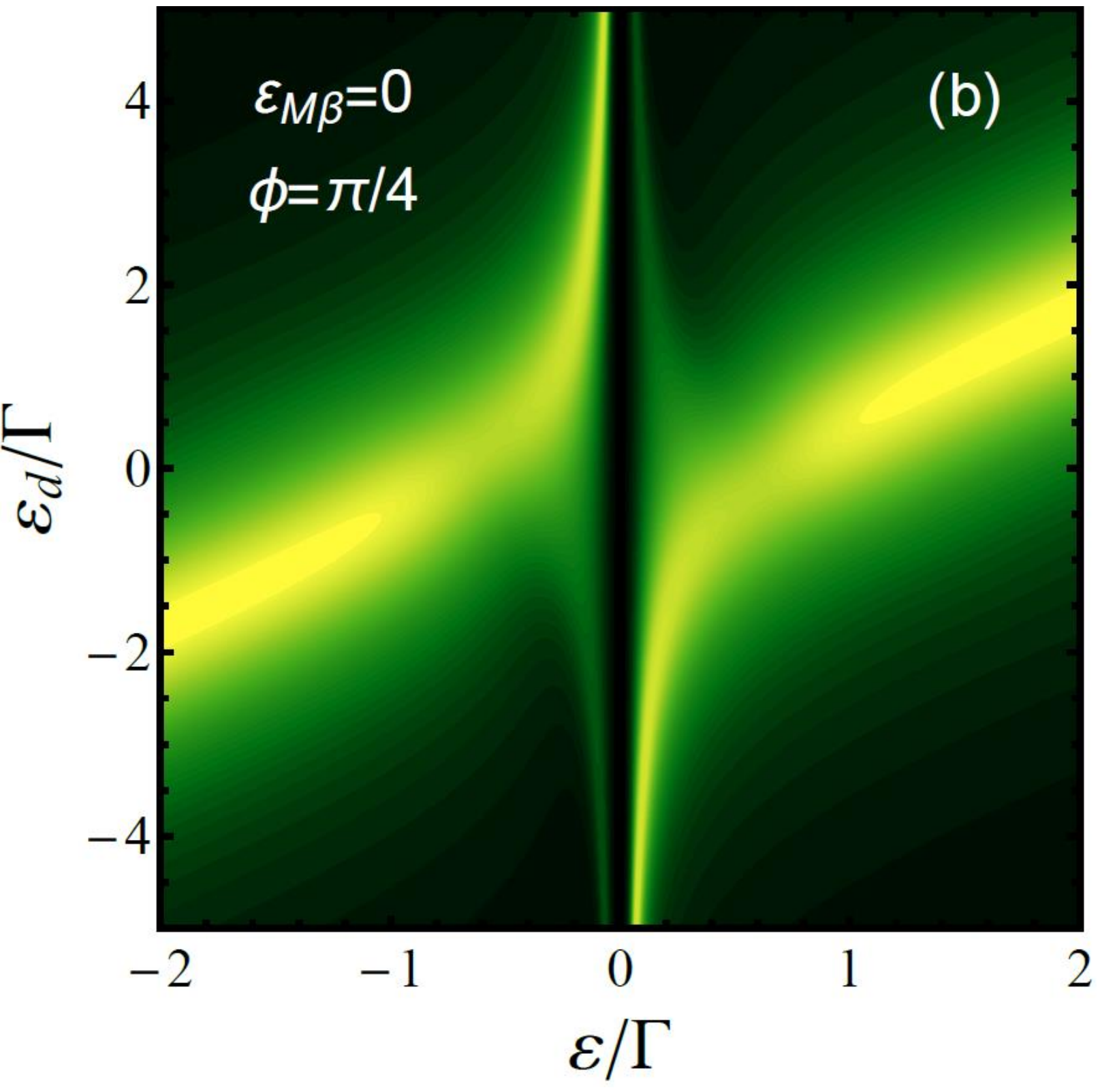}\includegraphics[width=0.33\textwidth]{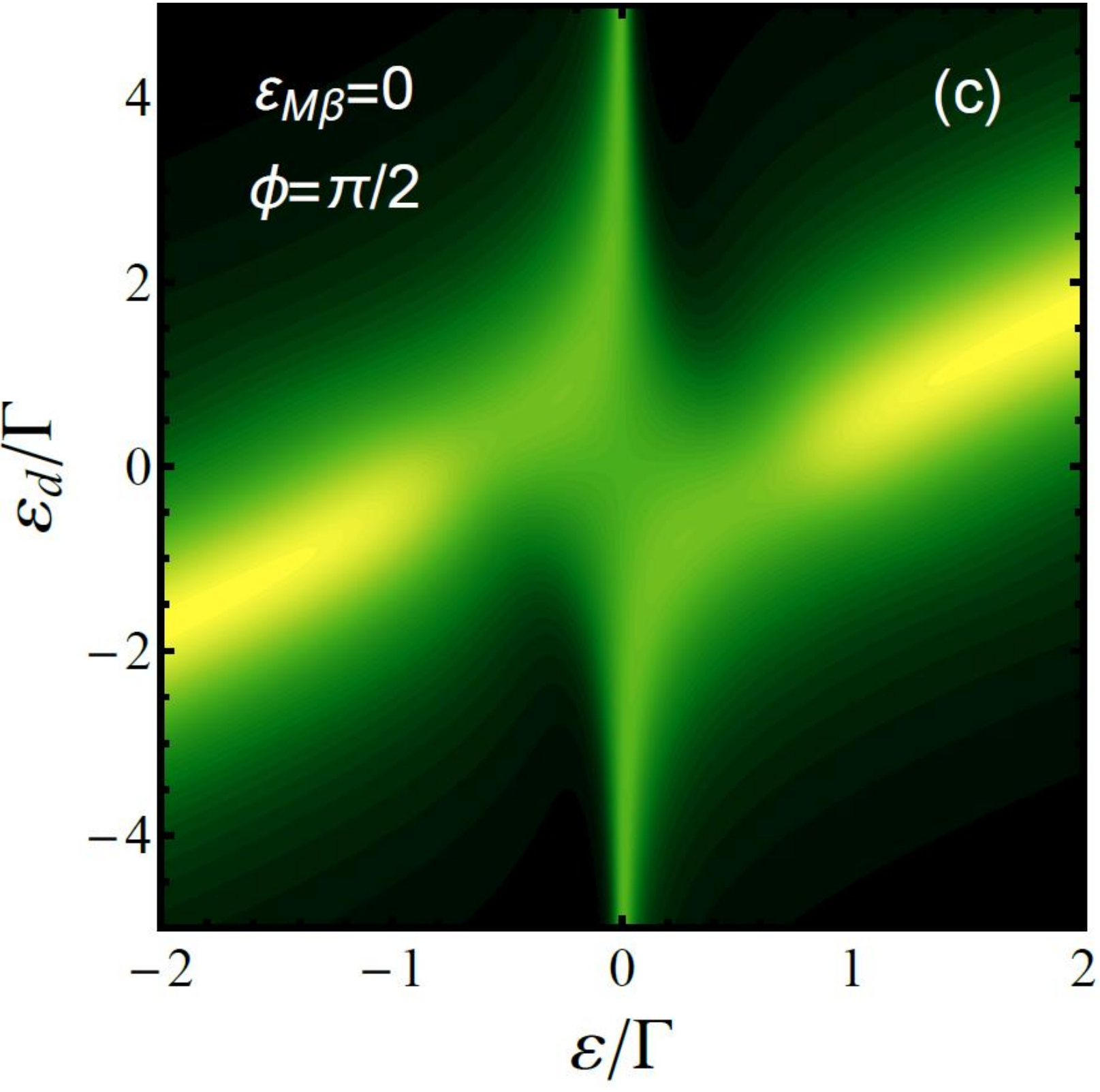}
\caption{Transmission probability maps as a function of energy and $\ve_{d}$ with $\alpha=0$. The AB flux phase in each panel is different: (a) $\phi/\pi=0$, (b) $\phi = \pi/4$, and (c)  $\phi = \pi/2$, respectively.}
\label{fig4}
\end{figure*}

\begin{figure*}[tbph]
\centering
\includegraphics[width=0.15\textwidth]{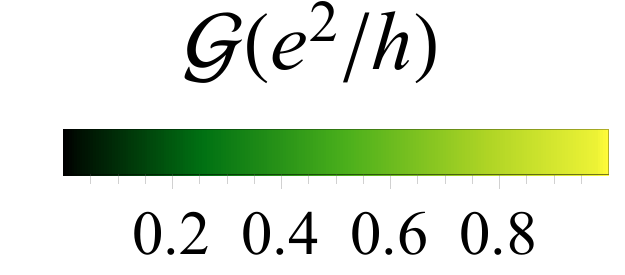}\\
\includegraphics[width=0.33\textwidth]{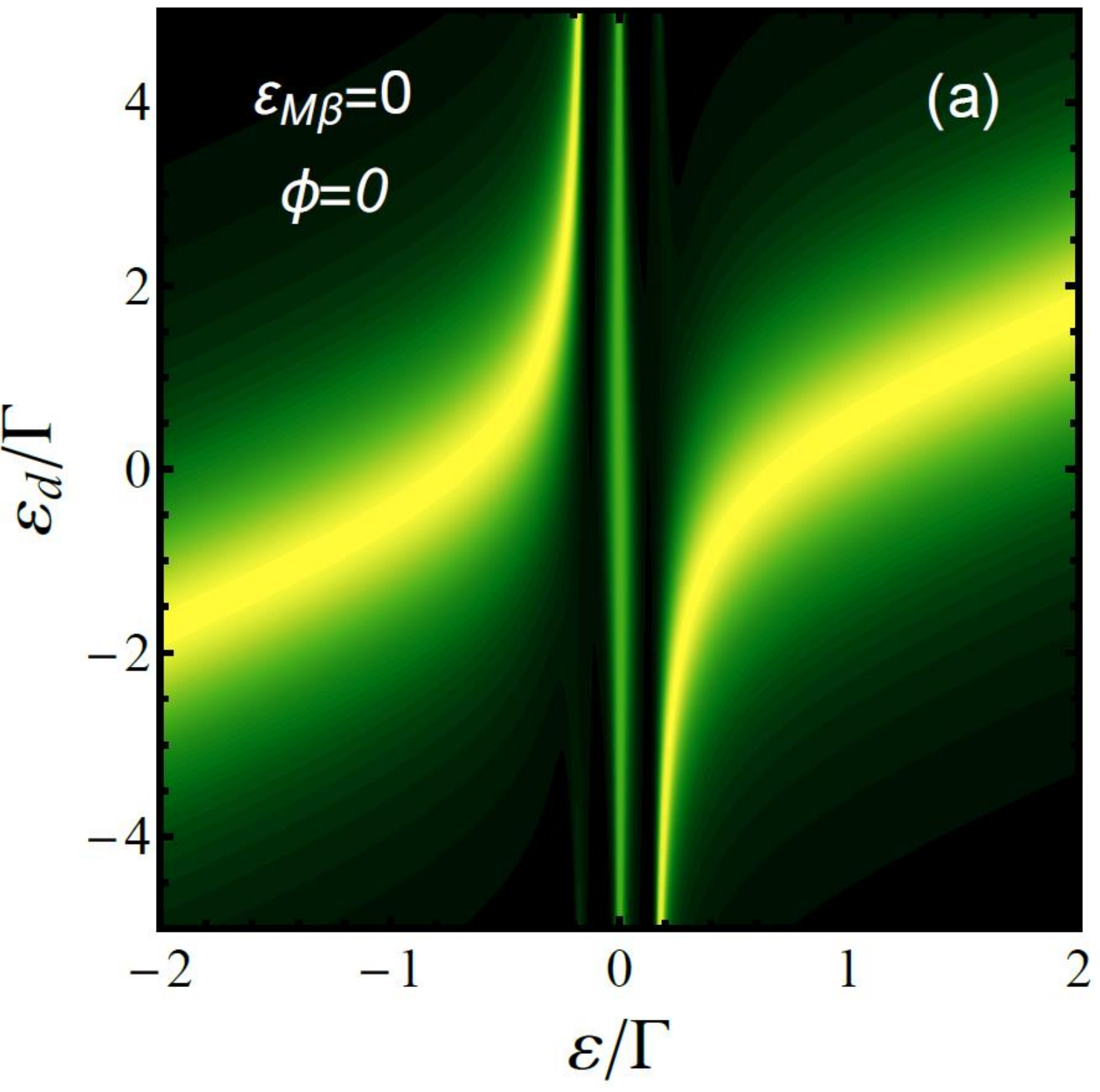}\includegraphics[width=0.33\textwidth]{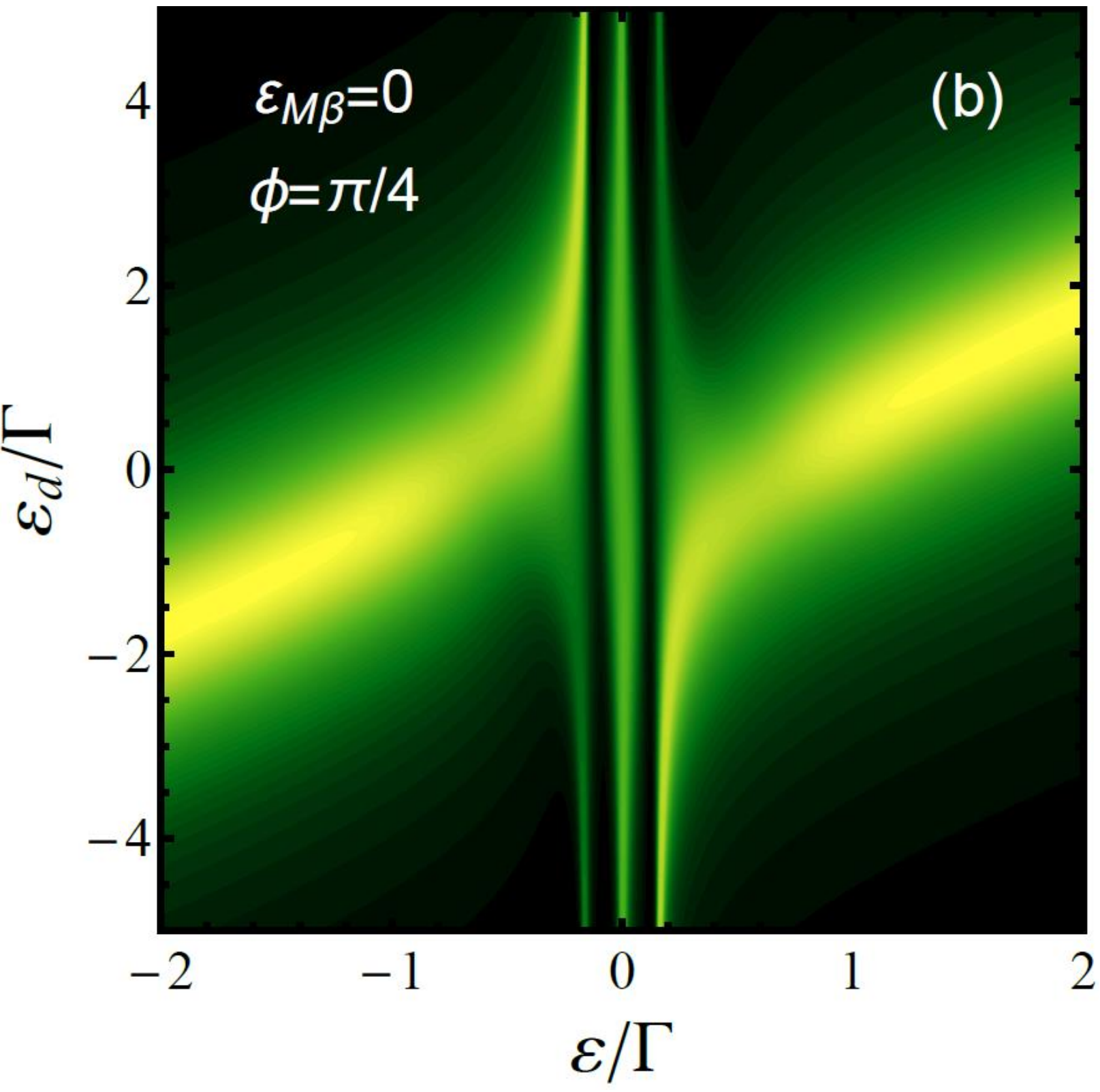}\includegraphics[width=0.33\textwidth]{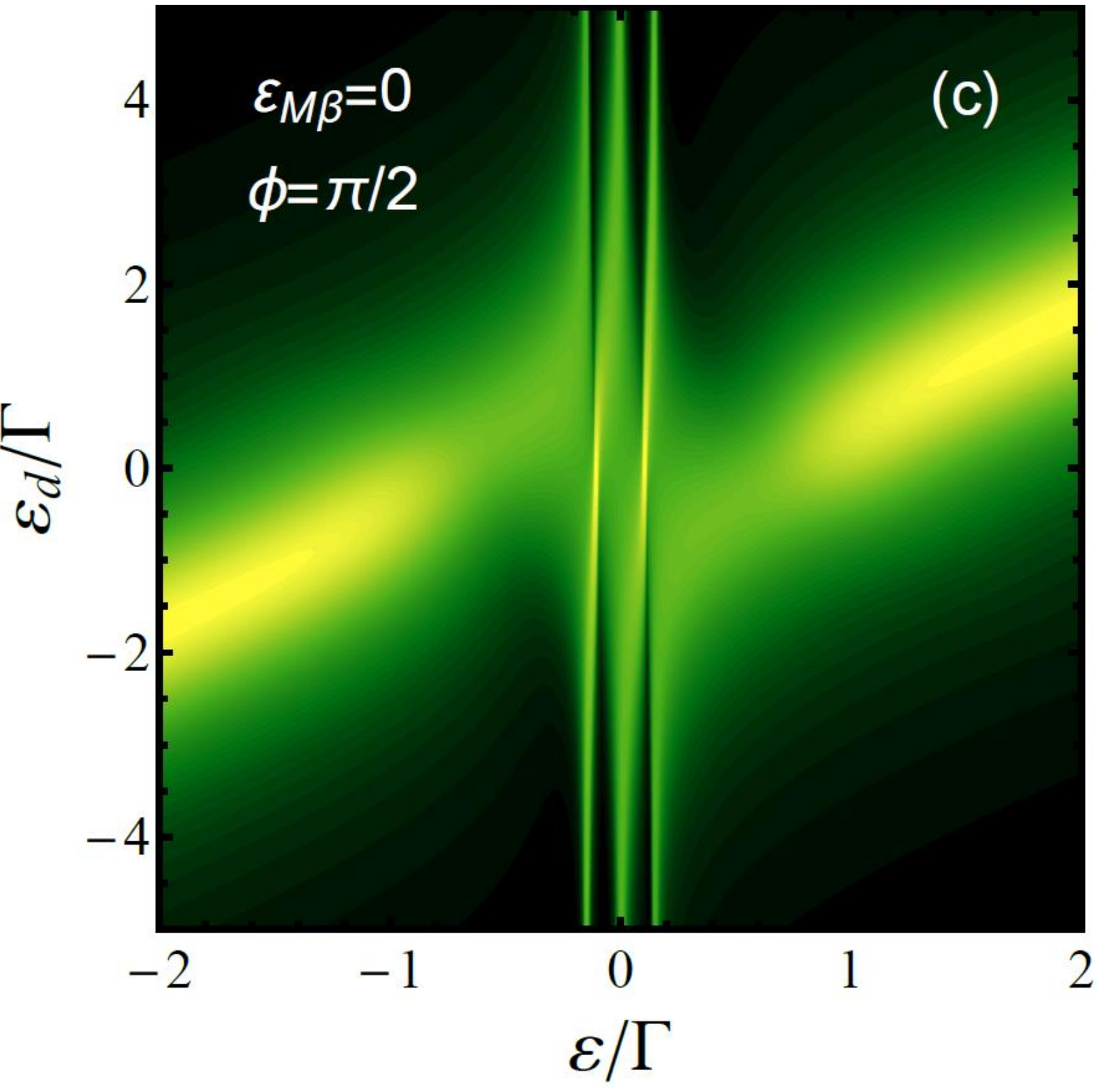}%\includegraphics[width=0.065\textwidth]{fig4bar.pdf}
\caption{Transmission probability maps as a function of energy and $\ve_{d}$ for interacting TNWs with $\alpha=0.15\,\Gamma$. AB fluxes are (a) $\phi/\pi=0$, (b) $\phi = \pi/4$, and (c)  $\phi = \pi/2$, respectively. Notice robustness of half-maximum peak at $\ve=0$ for all phase values, reflecting the interaction of MBSs across TNWs in the system.}
\label{fig5}
\end{figure*}

We show results for the interesting case that considers each TNW length sufficiently long to have vanishing coupling between the two MBSs belonging to the same nanowire, \emph{i.e.} $\ve_{M\beta}\rightarrow0$. In Figs.\ \ref{fig2} and \ref{fig3} we show the transmission probability and conductance across the QD as a function of the energy and Aharonov-Bohm phase, respectively. We start by setting $\alpha=0$ in Fig.\ \ref{fig2} panels (a)-(b) and in Fig.\ \ref{fig3}(a), which means that the MBSs belonging to different TNWs are disconnected. Figure \ref{fig2}(a) shows there is a phase-dependent behavior of the conductance, as three different phase values show rather distinct energy dependence. Only for $\phi=\pi/2$ (dark green dashed-dotted line) is the transmission probability $\mathcal{T}(\ve)\neq 0$ at $\ve=0$. This is the signature of the presence of MBS in the system, a half-maximum in conductance at $\ve=0$, ($\mathcal{G}(\ve=0)=e^2/2h$).\cite{Liu2011} In this case, the feature is not restricted to the zero-energy point, as the transmission exhibits a rather flat region around it. For other phase values (black solid and red dashed line), the conductance, in fact, dips to zero with a width similar to the plateau for $\phi=\pi/2$. In Fig.\ \ref{fig2}(b), the conductance curve for $\ve=0$ as a function of phase, supports the uniqueness of the $\phi=\pi/2$ phase value. In Fig.\ \ref{fig3}(a) we observe the Majorana signature, and the narrow flat region around zero energy, for every phase value $\phi_{m}=m\pi/2$, with $m$ an odd integer. The conductance curves illustrate the importance of interference of electronic paths through the QD and through the TNW. It is clear that even in the absence of direct MBS hybridization (since $\ve_{M\beta}=0$), the interferometer works and the varying flux is able to affect the conductance through the attached QD. The intrinsic entangled character of the MBSs is responsible for the effective electronic transfer across the TNW,\cite{Lu2012} while imparting an overall phase shift that cancels the conductance when the QD and the MBS are resonant at the Fermi level. This behavior underscores the possibility of implementing a flux-controlled operation in quantum computing architectures employing TNWs.\cite{Liu2011,Flensberg2011,Hyart2013}

In Fig.\ \ref{fig2}(c)-(d) and\ \ref{fig3}(b) we consider coupled TNWs with a fixed $\alpha=0.15\,\Gamma$ to study the effect of interacting MBSs across the TNWs. Figure \ref{fig2}(c) shows a phase-independent behavior in the conductance for $\ve=0$, the half-maximum conductance signature of MBS in the system extends for all phase values, as confirmed in Fig.\ \ref{fig2}(d). Moreover, the $\phi=\pi/2$ curve in this case (dark green dashed-dotted line) exhibits two symmetric sharp full conductance peaks located on both sides of the Fermi energy (at $\ve \sim \pm 0.1\,\Gamma$ here). Figure \ref{fig3}(b) shows that the peaks vanish quickly as the phase moves away from $\phi_{m}$. Besides, the flat plateau seen in the isolated interferometer case persists, although narrower than before.
The changes in conductance are remarkable, as it detects the connection between MBSs in the two TNWs, showing a half-maximum at zero energy, unaffected by phase changes, $\mathcal{G}(\ve=0,\phi)=e^2/2h$.  Such robustness with flux would be fairly evident and easy to identify in experiments.

Interestingly, these results are robust against changes in the dot energy level $\ve_{d}$, as shown in Figs.\ \ref{fig4} and \ref{fig5}. From Fig.\ \ref{fig4}, where $\alpha$ vanishes, we observe in panels (a) and (b) that the regular single-particle resonance behavior is apparent
when $\ve_d$ is away from the Fermi energy, and it is substantially independent of the AB phase.  However, for $\ve_d \simeq 0$, one sees a strong
conductance suppression whenever $\phi \neq \phi_m$. In panel (c), the MBS signature of half conductance maximum remains for $\phi =\pi/2$, even when the dot energy level moves away from the Fermi level ($\ve_{d}\neq \ve_F$).
The strong MBS interference reflects the leakage of the interferometer MBSs into the probing QD, as reported in the literature.\cite{Vernek2014}

In Fig.\ \ref{fig5}, where the TNW interaction $\alpha$ is turned on, the conductance exhibits a half-maximum value, essentially independent of the value of both the dot energy level and the AB phase, $\mathcal{G}(\ve=0,\phi,\ve_{d})=e^2/2h$. This result can be seen as the external MBS leaking into the QD through the coupling with the interferometer-TNW, substantially affecting the interferometer MBSs leakage.

\section{Summary}\label{secsummary}

We have studied the transport properties across a QD embedded between two normal leads used as a probe of MBS interactions across nearby TNWs. The QD is coupled to a TNW forming an Aharonov-Bohm interferometer, while a second TNW is placed nearby. The low-temperature conductance through the QD
changes drastically with the coupling between TNWs, $\alpha$.
When $\alpha=0$, the interferometer produces a half-maximum conductance only for the AB phase $\phi=\pi/2$ (and odd multiples), as reported before.\cite{Liu2011}
When $\alpha\neq0$, the neighboring TNWs are connected via separate MBSs, one of them belonging to the TNW in the interferometer setup. The low-temperature conductance exhibits an unusual Majorana signature (half-maximum value), {\em regardless} of the AB phase through the interferometer and the QD $\ve_{d}$ value. Hence, the conductance through the QD sensitively reflects the presence/interaction between MBSs, and such behavior is robust to changes in other system parameters.
These results suggest that such behavior could be used to probe the connection between MBSs located in different TNWs.  Setting the phase away from $\pi/2$ and at low temperature (so that the transmission at zero-energy dominates), the conductance will show a value $\mathcal{G}(\phi\neq\phi_{m},\ve_{d})=0$ if the MBSs are uncoupled.  In contrast, if the MBSs are connected, the conductance measurement at low temperature
will yield $\mathcal{G}(\phi\neq\phi_{m},\ve_{d}) \simeq e^2/2h$, independently of the position of the QD energy level.

\begin{acknowledgements}
J.P.R.-A. is grateful for the hospitality at Ohio University and the support from scholarship CONICYT-Chile No.\ 21141034. P.A.O. acknowledges support from FONDECYT Grant No.\ 1140571, and S.E.U. acknowledges support from NSF Grant No.\ DMR 1508325.
\end{acknowledgements}

\appendix

\section{Quantum dot Green's function}

In this appendix, we detail the procedure used to obtain the QD (retarded) Green's function (GF) using the equation of motion method. It is important to mention that we avoid the superscript $r$ in the retarded GF, in order to simplify the notation (e.g. Eq.\ (\ref{ABenergy})).
We use Eq.\ (\ref{ABenergy}) with the Hamiltonian in Eq.\ (\ref{H0}), with regular fermionic terms given by Eqs.\ (\ref{Hleads}-\ref{Hleadsdot}) while the Majorana contribution is taken in the form of Eq.\ (\ref{HMBSconf}). The dot GF reads (with $\ve \rightarrow \ve + i0^+$),
\begin{equation}
\ve\la\la d,d^{\dag}\ra\ra_{\ve}=\la\{d,d^{\dag}\}\ra+\la\la[d,H],d^{\dag}\ra\ra_{\ve}\,. \label{dddaga}
\end{equation}
Then, after taking into account the commutator, we have
\begin{eqnarray}
(\ve-\ve_{d})\la\la d,d^{\dag}\ra\ra_{\ve}=1&+&\sum_{j,k}t_{j}^{\ast}\la\la c_{j,k},d^{\dag}\ra\ra_{\ve}\\ \nonumber
&+&\left(\frac{\lambda_{2}^{\ast}-\lambda_{1}^{\ast}}{\sqrt{2}}\right)\la\la f_{\gamma},d^{\dag}\ra\ra_{\ve}\\ \nonumber
&-&\left(\frac{\lambda_{2}^{\ast}+\lambda_{1}^{\ast}}{\sqrt{2}}\right)\la\la f_{\gamma}^{\dag},d^{\dag}\ra\ra_{\ve}\,, \label{dddaga2}
\end{eqnarray}
where we note additional GFs that must be calculated. This is a recursive procedure that yields a closed set of equations in this non-interacting problem without further approximations. Repeating the process for the second term on the right-hand side of Eq.\ (\ref{dddaga2}), we obtain
\begin{equation}
\la\la c_{j,k},d^{\dag}\ra\ra_{\ve}=\frac{t_{j}}{\ve-\ve_{k}}\la\la d,d^{\dag}\ra\ra_{\ve}\,. \label{cd}
\end{equation}

This function represents the coupling between the leads and the QD.\@  The self-energy due to the leads is then
\begin{equation}
\sum_{j,k}\frac{|t_{j}|^{2}}{\ve-\ve_{k}}\equiv\left(\Sigma_{L}(\ve)+\Sigma_{R}(\ve)\right)\,.
\end{equation}
The wide band approximation yields an energy-independent self-energy $\Sigma_{j}(\ve)=-i\Gamma_{j}$. We further adopt symmetric lead coupling, fixing $\Gamma_{L}=\Gamma_{R}\equiv\Gamma/2$, so that we re-write Eq.\ (\ref{dddaga2}) as
\begin{eqnarray}
(\ve-\ve_{d}+&i&\Gamma)\la\la d,d^{\dag}\ra\ra_{\ve}=\\ \nonumber
&1&+\left(\frac{\lambda_{2}^{\ast}-\lambda_{1}^{\ast}}{\sqrt{2}}\right)\la\la f_{\gamma},d^{\dag}\ra\ra_{\ve}-\left(\frac{\lambda_{2}^{\ast}+\lambda_{1}^{\ast}}{\sqrt{2}}\right)\la\la f_{\gamma}^{\dag},d^{\dag}\ra\ra_{\ve}\,. \label{ddd}
\end{eqnarray}

The GFs involving the interferometer MBSs are given by
\begin{widetext}
\begin{eqnarray}
(\ve-\ve_{M\gamma})\la\la f_{\gamma},d^{\dag}\ra\ra_{\ve}&=&\left(\frac{\lambda_{2}-\lambda_{1}}{\sqrt{2}}\right)\la\la d,d^{\dag}\ra\ra_{\ve}+\left(\frac{\lambda_{2}^{\ast}+\lambda_{1}^{\ast}}{\sqrt{2}}\right)\la\la d^{\dag},d^{\dag}\ra\ra_{\ve}-\frac{\alpha}{2}\left(\la\la f_{\eta},d^{\dag}\ra\ra_{\ve}+\la\la f_{\eta}^{\dag},d^{\dag}\ra\ra_{\ve}\right)\,, \label{ggamma} \\
(\ve+\ve_{M\gamma})\la\la f_{\gamma}^{\dag},d^{\dag}\ra\ra_{\ve}&=&-\left(\frac{\lambda_{2}+\lambda_{1}}{\sqrt{2}}\right)\la\la d,d^{\dag}\ra\ra_{\ve}-\left(\frac{\lambda_{2}^{\ast}-\lambda_{1}^{\ast}}{\sqrt{2}}\right)\la\la d^{\dag},d^{\dag}\ra\ra_{\ve}+\frac{\alpha}{2}\left(\la\la f_{\eta},d^{\dag}\ra\ra_{\ve}+\la\la f_{\eta}^{\dag},d^{\dag}\ra\ra_{\ve}\right)\,. \label{ggamma2}
\end{eqnarray}
\end{widetext}

The calculation requires three more GFs, including an anomalous dot GF, and those corresponding to the external MBSs. The last two are given by
\begin{eqnarray}
(\ve-\ve_{M\eta})\la\la f_{\eta},d^{\dag}\ra\ra_{\ve}&=&\\ \nonumber
&-&\frac{\alpha}{2}\left(\la\la f_{\gamma},d^{\dag}\ra\ra_{\ve}-\la\la f_{\gamma}^{\dag},d^{\dag}\ra\ra_{\ve}\right)\,, \label{geta}\\
(\ve+\ve_{M\eta})\la\la f_{\eta}^{\dag},d^{\dag}\ra\ra_{\ve}&=&\\ \nonumber
&-&\frac{\alpha}{2}\left(\la\la f_{\gamma},d^{\dag}\ra\ra_{\ve}-\la\la f_{\gamma}^{\dag},d^{\dag}\ra\ra_{\ve}\right)\,. \label{geta2}
\end{eqnarray}
The anomalous term is given by
\begin{eqnarray}
(\ve&+&\ve_{d})\la\la d^{\dag},d^{\dag}\ra\ra_{\ve}=-\sum_{j,k}t_{j}\la\la c_{j,k}^{\dag},d^{\dag}\ra\ra_{\ve}\\ \nonumber
&+&\left(\frac{\lambda_{2}+\lambda_{1}}{\sqrt{2}}\right)\la\la f_{\gamma},d^{\dag}\ra\ra_{\ve}-\left(\frac{\lambda_{2}-\lambda_{1}}{\sqrt{2}}\right)\la\la f_{\gamma}^{\dag},d^{\dag}\ra\ra_{\ve}\,,\label{drara}
\end{eqnarray}
where we note another GF involving the leads. Assuming electron-hole symmetry in the system, we can carry out the same analysis, which yields
\begin{equation}
(\ve+\ve_{k})\la\la c_{j,k}^{\dag},d^{\dag}\ra\ra_{\ve}=-t_{j}^{\ast}\la\la d^{\dag},d^{\dag}\ra\ra_{\ve}\,,
\end{equation}
and
\begin{equation}
\sum_{j,k}\frac{|t_{j}|^{2}}{\ve+\ve_{k}}=-i\Gamma\,.
\end{equation}
Replacing it in Eq.\ (\ref{drara}) we obtain
\begin{eqnarray}
(\ve+\ve_{d}&+&i\Gamma)\la\la d^{\dag},d^{\dag}\ra\ra_{\ve}=\\ \nonumber
&&\left(\frac{\lambda_{2}+\lambda_{1}}{\sqrt{2}}\right)\la\la f_{\gamma},d^{\dag}\ra\ra_{\ve}-\left(\frac{\lambda_{2}-\lambda_{1}}{\sqrt{2}}\right)\la\la f_{\gamma}^{\dag},d^{\dag}\ra\ra_{\ve}\,. \label{dddd}
\end{eqnarray}
We have now achieved a complete set of equations. Replacing Eqs.\ (\ref{geta}-\ref{drara}) into Eqs.\ (\ref{ggamma}-\ref{ggamma2}), we write
\begin{widetext}
\begin{eqnarray}
\left[\ve-\ve_{M\gamma}-\frac{\alpha^{2}}{2}K_{\eta}(\ve)\left(1+\frac{\alpha^{2}}{2}J_{\eta\gamma}^{(+)}(\ve)\right)\right]\la\la f_{\gamma},d^{\dag}\ra\ra_{\ve}&=&\left[\left(\frac{\lambda_{2}-\lambda_{1}}{\sqrt{2}}\right)+\left(\frac{\lambda_{2}+\lambda_{1}}{\sqrt{2}}\right)\frac{\alpha^{2}}{2}J_{\eta\gamma}^{(+)}(\ve)\right]\la\la d,d^{\dag}\ra\ra_{\ve} \nonumber \\
&+&\left[\left(\frac{\lambda_{2}^{\ast}+\lambda_{1}^{\ast}}{\sqrt{2}}\right)+\left(\frac{\lambda_{2}^{\ast}-\lambda_{1}^{\ast}}{\sqrt{2}}\right)\frac{\alpha^{2}}{2}J_{\eta\gamma}^{(+)}(\ve)\right]\la\la d^{\dag},d^{\dag}\ra\ra_{\ve}\,, \label{fddaga}
\end{eqnarray}
\begin{eqnarray}
\left[\ve+\ve_{M\gamma}-\frac{\alpha^{2}}{2}K_{\eta}(\ve)\left(1+\frac{\alpha^{2}}{2}J_{\eta\gamma}^{(-)}(\ve)\right)\right]\la\la f_{\gamma}^{\dag},d^{\dag}\ra\ra_{\ve}&=&-\left[\left(\frac{\lambda_{2}+\lambda_{1}}{\sqrt{2}}\right)+\left(\frac{\lambda_{2}-\lambda_{1}}{\sqrt{2}}\right)\frac{\alpha^{2}}{2}J_{\eta\gamma}^{(-)}(\ve)\right]\la\la d,d^{\dag}\ra\ra_{\ve} \nonumber \\
&-&\left[\left(\frac{\lambda_{2}^{\ast}-\lambda_{1}^{\ast}}{\sqrt{2}}\right)+\left(\frac{\lambda_{2}^{\ast}+\lambda_{1}^{\ast}}{\sqrt{2}}\right)\frac{\alpha^{2}}{2}J_{\eta\gamma}^{(-)}(\ve)\right]\la\la d^{\dag},d^{\dag}\ra\ra_{\ve}\,, \label{fdagaddaga}
\end{eqnarray}
\end{widetext}
where we have defined the functions
\begin{equation}
K_{\eta}(\ve)=\frac{\ve}{\ve^{2}-\ve_{M\eta}^{2}}\,, \label{keta}
\end{equation}
\begin{eqnarray}
J_{\eta\gamma}^{(-)}(\ve)&=&\frac{K_{\eta}(\ve)}{\ve-\ve_{M\gamma}-(\alpha^{2}/2)K_{\eta}(\ve)}\,, \\
J_{\eta\gamma}^{(+)}(\ve)&=&\frac{K_{\eta}(\ve)}{\ve+\ve_{M\gamma}-(\alpha^{2}/2)K_{\eta}(\ve)}\,. \label{jotas}
\end{eqnarray}

Finally, replacing Eqs.\ (\ref{fddaga}-\ref{fdagaddaga}) into Eq.\ (\ref{ddd}) we arrive to the QD GF.

As in the main text, we define $\lambda_{2}=\lambda_{2}^{\ast}=\lambda$ and $\lambda_{1}=\lambda e^{i\phi}$. We also fix $\ve_{\beta}=0$, so that $K_{\eta}(\ve)=1/\ve$ and $J_{\eta\gamma}^{(+)}(\ve)=J_{\eta\gamma}^{(-)}(\ve)=J(\ve)$. Defining the function
\begin{equation}
\Omega(\ve)=\frac{1}{2\ve}\left(1+\frac{\alpha^{2}}{2}J(\ve)\right)\,,
\end{equation}
the QD Green function is given by
\begin{widetext}
\begin{equation}
\left[\la\la d,d^{\dag}\ra\ra_{\ve}\right]^{-1} \equiv  F_2 (\ve) + i \Gamma F_1 (\ve) \, ,
\end{equation}
where
\begin{equation}
F_{1}(\ve)= 1+\frac{4\lambda^{4}(\sin^{2}(\phi)+(\alpha^{4}/4)J^{2}(\ve)\cos^{2}(\phi))}{((\ve-\alpha^{2}\Omega(\ve))(\ve+\ve_{d})-2\lambda^{2})^{2}+(\ve-\alpha^{2}\Omega(\ve))^{2}\Gamma^{2}} \,,
\end{equation}
\begin{equation}
F_{2}(\ve)=\ve-\ve_{d}-\frac{2\lambda^{2}}{\ve-\alpha^{2}\Omega(\ve)}\left(1+\frac{2\lambda^{2}(\sin^{2}(\phi)+(\alpha^{4}/4)J^{2}(\ve)\cos^{2}(\phi))((\ve-\alpha^{2}\Omega(\ve))(\ve+\ve_{d})-2\lambda^{2})}{((\ve-\alpha^{2}\Omega(\ve))(\ve+\ve_{d})-2\lambda^{2})^{2}+(\ve-\alpha^{2}\Omega(\ve))^{2}\Gamma^{2}}\right) \,.
\end{equation}
\end{widetext}
We explore the behavior of $\mathcal{T}(\ve \rightarrow 0) = -\Gamma \, {\rm Im} \, \la\la d,d^{\dag}\ra\ra_{\ve}$ for fixed $\ve_{d}=0$. For the isolated interferometer case ($\alpha=0$) we have
\begin{eqnarray}
\lim_{\ve\rightarrow 0}F_{1}(\ve)&=&1+\sin^{2}(\phi)\,,\\
\lim_{\ve\rightarrow 0}F_{2}(\ve)&=&\left\{
  \begin{array}{ll}
    0, & \hbox{if $\phi=\pi/2$;} \\
    -\infty, & \hbox{if $\phi\neq\pi/2$.}
  \end{array}
\right.
\end{eqnarray}
This yields
\begin{equation}
\lim_{\ve\rightarrow 0}\mathcal{T}(\ve)=\left\{
  \begin{array}{ll}
    1/2, & \hbox{if $\phi=\pi/2$;} \\
    0, & \hbox{if $\phi\neq\pi/2$.}
  \end{array}
\right.
\end{equation}
On the other hand, for $\alpha\neq 0$, we have
\begin{eqnarray}
\lim_{\ve\rightarrow 0}F_{1}(\ve)&=&2\,,\\
\lim_{\ve\rightarrow 0}F_{2}(\ve)&=&0\,,
\end{eqnarray}
regardless of the phase value, so that
\begin{equation}
\lim_{\ve\rightarrow 0}\mathcal{T}(\ve)=1/2\,.
\end{equation}
This limiting behavior is reflected in the plots shown in the main text.

Lastly, we present the eigenvalues of the Hamiltonian $H_{\text{dot}}+H_{\text{MBS}}$, obtained by direct diagonalization. These are
closely related to the poles in the QD GFs, and therefore to the resonances in $\mathcal{T}(\ve)$. They are particle-hole symmetric, as expected, and
given by $\ve_i^{\pm} = \pm \ve_i$, with
\begin{widetext}
\begin{eqnarray}
\ve_{0} &=& 0\,, \\
\sqrt{2}\,\ve_{1} &=& \sqrt{\ve_{d}^{2}+4\lambda^{2}+\alpha^{2}+\sqrt{\ve_{d}^{2}(\ve_{d}^{2}+8\lambda^{2}-2\alpha^{2})+\alpha^{4}+16\lambda^{4}\sin^{2}(\phi)}}\,, \\
\sqrt{2}\,\ve_{2} &=& \sqrt{\ve_{d}^{2}+4\lambda^{2}+\alpha^{2}-\sqrt{\ve_{d}^{2}(\ve_{d}^{2}+8\lambda^{2}-2\alpha^{2})+\alpha^{4}+16\lambda^{4}\sin^{2}(\phi)}}\,.
\end{eqnarray}
\end{widetext}

\begin{figure}[tbph]
\centering
\includegraphics[width=0.45\textwidth]{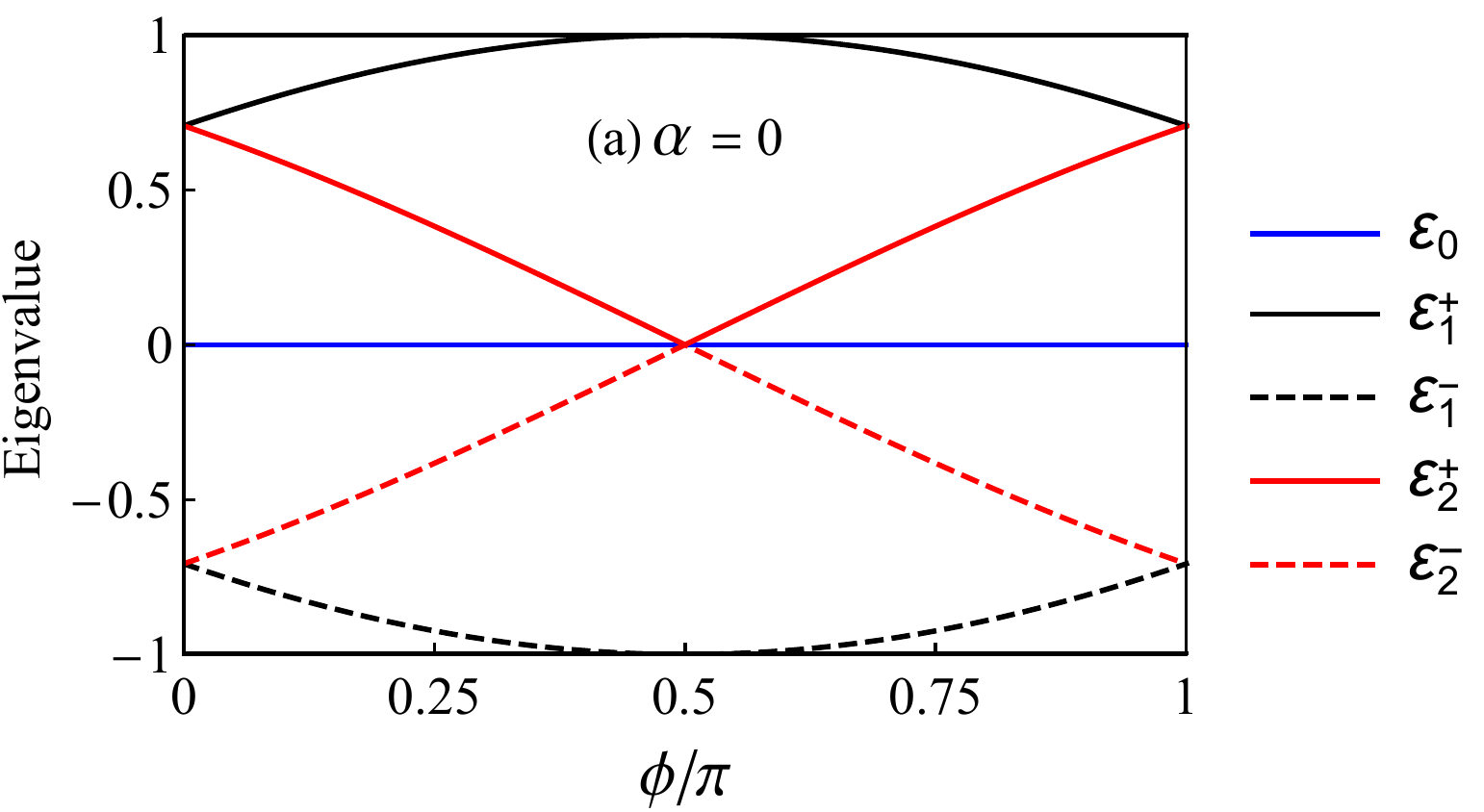}\\
\includegraphics[width=0.45\textwidth]{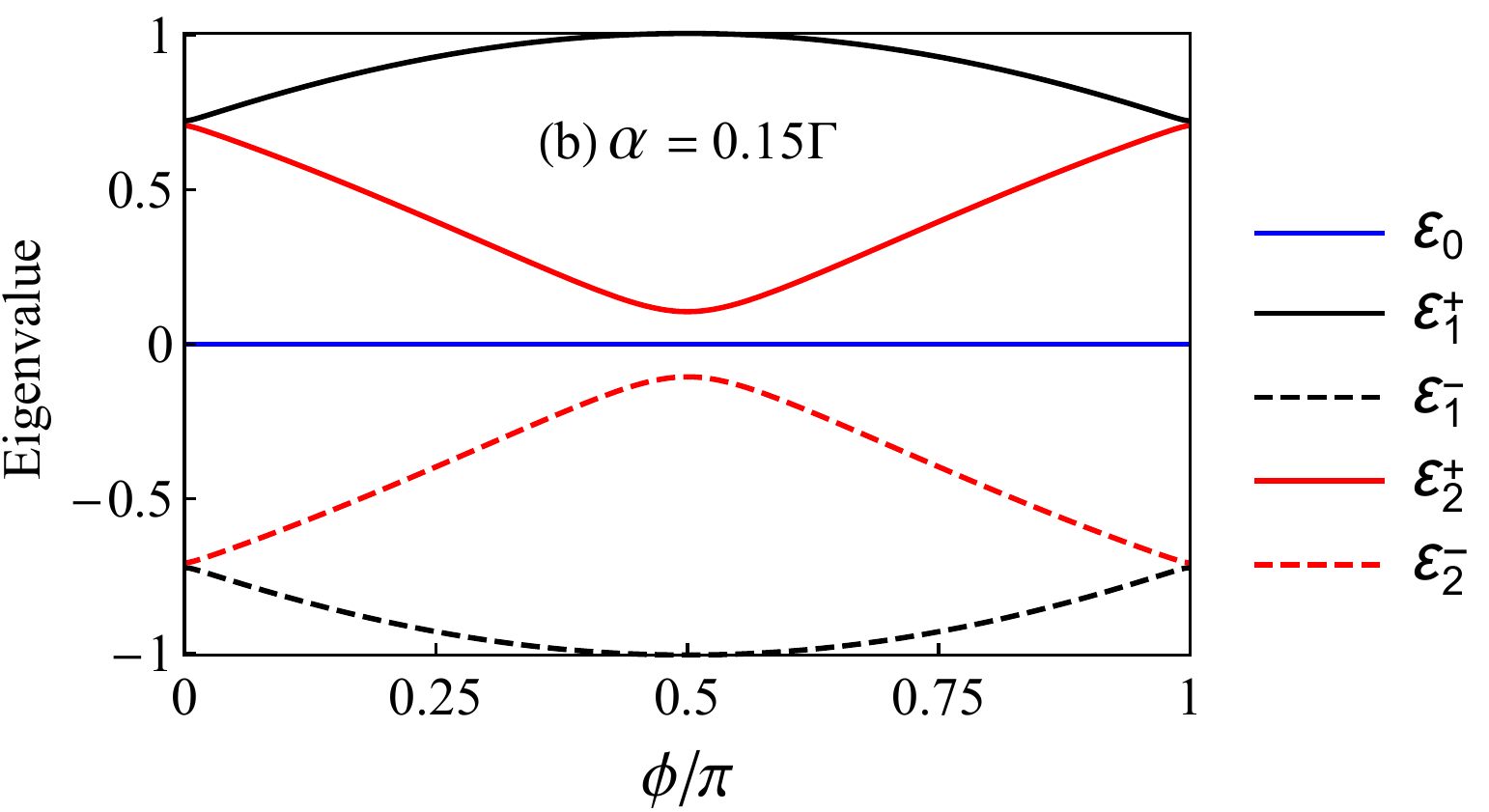}
\caption{Eigenvalues of $H_{\text{dot}}+H_{\text{MBS}}$ as function of AB phase for fixed $\ve_{d}=0$. (a) $\alpha=0$ and (b) $\alpha=0.15\,\Gamma$.  Eigenvalues in units of $\Gamma$.}
\label{fig6}
\end{figure}

As shown in Fig.\ \ref{fig6}, it is clear that the half-maximum conductance for decoupled TNWs occurs at the triple degeneracy point for $\phi = \pi/2$. \cite{Liu2011}  For $\alpha \neq 0$, in contrast, the degeneracy is split, and the mode at zero results in a half-conductance for any value of $\phi$.

\noindent

\bibliographystyle{apsrev4-1}
\bibliography{biblio}

\begin{thebibliography}{39}%
\makeatletter
\providecommand \@ifxundefined [1]{%
 \@ifx{#1\undefined}
}%
\providecommand \@ifnum [1]{%
 \ifnum #1\expandafter \@firstoftwo
 \else \expandafter \@secondoftwo
 \fi
}%
\providecommand \@ifx [1]{%
 \ifx #1\expandafter \@firstoftwo
 \else \expandafter \@secondoftwo
 \fi
}%
\providecommand \natexlab [1]{#1}%
\providecommand \enquote  [1]{``#1''}%
\providecommand \bibnamefont  [1]{#1}%
\providecommand \bibfnamefont [1]{#1}%
\providecommand \citenamefont [1]{#1}%
\providecommand \href@noop [0]{\@secondoftwo}%
\providecommand \href [0]{\begingroup \@sanitize@url \@href}%
\providecommand \@href[1]{\@@startlink{#1}\@@href}%
\providecommand \@@href[1]{\endgroup#1\@@endlink}%
\providecommand \@sanitize@url [0]{\catcode `\\12\catcode `\$12\catcode
  `\&12\catcode `\#12\catcode `\^12\catcode `\_12\catcode `\%12\relax}%
\providecommand \@@startlink[1]{}%
\providecommand \@@endlink[0]{}%
\providecommand \url  [0]{\begingroup\@sanitize@url \@url }%
\providecommand \@url [1]{\endgroup\@href {#1}{\urlprefix }}%
\providecommand \urlprefix  [0]{URL }%
\providecommand \Eprint [0]{\href }%
\providecommand \doibase [0]{http://dx.doi.org/}%
\providecommand \selectlanguage [0]{\@gobble}%
\providecommand \bibinfo  [0]{\@secondoftwo}%
\providecommand \bibfield  [0]{\@secondoftwo}%
\providecommand \translation [1]{[#1]}%
\providecommand \BibitemOpen [0]{}%
\providecommand \bibitemStop [0]{}%
\providecommand \bibitemNoStop [0]{.\EOS\space}%
\providecommand \EOS [0]{\spacefactor3000\relax}%
\providecommand \BibitemShut  [1]{\csname bibitem#1\endcsname}%
\let\auto@bib@innerbib\@empty
%</preamble>
\bibitem [{\citenamefont {Majorana}(1937)}]{Majorana2008}%
  \BibitemOpen
  \bibfield  {author} {\bibinfo {author} {\bibfnamefont {E.}~\bibnamefont
  {Majorana}},\ }\href {\doibase 10.1007/BF02961314} {\bibfield  {journal}
  {\bibinfo  {journal} {Nuovo Cimento}\ }\textbf {\bibinfo {volume} {14}},\
  \bibinfo {pages} {171} (\bibinfo {year} {1937})}\BibitemShut {NoStop}%
\bibitem [{\citenamefont {Wilczek}(2009)}]{Wilczek2009}%
  \BibitemOpen
  \bibfield  {author} {\bibinfo {author} {\bibfnamefont {F.}~\bibnamefont
  {Wilczek}},\ }\href {\doibase 10.1038/nphys1380} {\bibfield  {journal}
  {\bibinfo  {journal} {Nat. Phys.}\ }\textbf {\bibinfo {volume} {5}},\
  \bibinfo {pages} {614} (\bibinfo {year} {2009})}\BibitemShut {NoStop}%
\bibitem [{\citenamefont {Alicea}(2016)}]{alicea2016}%
  \BibitemOpen
  \bibfield  {author} {\bibinfo {author} {\bibfnamefont {J.}~\bibnamefont
  {Alicea}},\ }\href
  {http://www.nature.com/nature/journal/v531/n7593/full/531177a.html}
  {\bibfield  {journal} {\bibinfo  {journal} {Nature}\ }\textbf {\bibinfo
  {volume} {531}},\ \bibinfo {pages} {177} (\bibinfo {year}
  {2016})}\BibitemShut {NoStop}%
\bibitem [{\citenamefont {Alicea}\ \emph {et~al.}(2011)\citenamefont {Alicea},
  \citenamefont {Oreg}, \citenamefont {Refael}, \citenamefont {Von~Oppen},\
  and\ \citenamefont {Fisher}}]{alicea2011}%
  \BibitemOpen
  \bibfield  {author} {\bibinfo {author} {\bibfnamefont {J.}~\bibnamefont
  {Alicea}}, \bibinfo {author} {\bibfnamefont {Y.}~\bibnamefont {Oreg}},
  \bibinfo {author} {\bibfnamefont {G.}~\bibnamefont {Refael}}, \bibinfo
  {author} {\bibfnamefont {F.}~\bibnamefont {Von~Oppen}}, \ and\ \bibinfo
  {author} {\bibfnamefont {M.~P.}\ \bibnamefont {Fisher}},\ }\href
  {http://www.nature.com/nphys/journal/v7/n5/full/nphys1915.html} {\bibfield
  {journal} {\bibinfo  {journal} {Nat. Phys.}\ }\textbf {\bibinfo {volume}
  {7}},\ \bibinfo {pages} {412} (\bibinfo {year} {2011})}\BibitemShut {NoStop}%
\bibitem [{\citenamefont {Fu}\ and\ \citenamefont {Kane}(2008)}]{Fu2008}%
  \BibitemOpen
  \bibfield  {author} {\bibinfo {author} {\bibfnamefont {L.}~\bibnamefont
  {Fu}}\ and\ \bibinfo {author} {\bibfnamefont {C.~L.}\ \bibnamefont {Kane}},\
  }\href {\doibase 10.1103/PhysRevLett.100.096407} {\bibfield  {journal}
  {\bibinfo  {journal} {Phys. Rev. Lett.}\ }\textbf {\bibinfo {volume} {100}},\
  \bibinfo {pages} {096407} (\bibinfo {year} {2008})}\BibitemShut {NoStop}%
\bibitem [{\citenamefont {Ivanov}(2001)}]{Ivanov2001}%
  \BibitemOpen
  \bibfield  {author} {\bibinfo {author} {\bibfnamefont {D.~A.}\ \bibnamefont
  {Ivanov}},\ }\href {\doibase 10.1103/PhysRevLett.86.268} {\bibfield
  {journal} {\bibinfo  {journal} {Phys. Rev. Lett.}\ }\textbf {\bibinfo
  {volume} {86}},\ \bibinfo {pages} {268} (\bibinfo {year} {2001})}\BibitemShut
  {NoStop}%
\bibitem [{\citenamefont {Lutchyn}\ \emph {et~al.}(2010)\citenamefont
  {Lutchyn}, \citenamefont {Sau},\ and\ \citenamefont
  {Das~Sarma}}]{Lutchyn2010}%
  \BibitemOpen
  \bibfield  {author} {\bibinfo {author} {\bibfnamefont {R.~M.}\ \bibnamefont
  {Lutchyn}}, \bibinfo {author} {\bibfnamefont {J.~D.}\ \bibnamefont {Sau}}, \
  and\ \bibinfo {author} {\bibfnamefont {S.}~\bibnamefont {Das~Sarma}},\ }\href
  {\doibase 10.1103/PhysRevLett.105.077001} {\bibfield  {journal} {\bibinfo
  {journal} {Phys. Rev. Lett.}\ }\textbf {\bibinfo {volume} {105}},\ \bibinfo
  {pages} {077001} (\bibinfo {year} {2010})}\BibitemShut {NoStop}%
\bibitem [{\citenamefont {Oreg}\ \emph {et~al.}(2010)\citenamefont {Oreg},
  \citenamefont {Refael},\ and\ \citenamefont {von Oppen}}]{Oreg2010}%
  \BibitemOpen
  \bibfield  {author} {\bibinfo {author} {\bibfnamefont {Y.}~\bibnamefont
  {Oreg}}, \bibinfo {author} {\bibfnamefont {G.}~\bibnamefont {Refael}}, \ and\
  \bibinfo {author} {\bibfnamefont {F.}~\bibnamefont {von Oppen}},\ }\href
  {\doibase 10.1103/PhysRevLett.105.177002} {\bibfield  {journal} {\bibinfo
  {journal} {Phys. Rev. Lett.}\ }\textbf {\bibinfo {volume} {105}},\ \bibinfo
  {pages} {177002} (\bibinfo {year} {2010})}\BibitemShut {NoStop}%
\bibitem [{\citenamefont {Nadj-Perge}\ \emph {et~al.}(2014)\citenamefont
  {Nadj-Perge}, \citenamefont {Drozdov}, \citenamefont {Li}, \citenamefont
  {Chen}, \citenamefont {Jeon}, \citenamefont {Seo}, \citenamefont {MacDonald},
  \citenamefont {Bernevig},\ and\ \citenamefont {Yazdani}}]{NadjPerge2014}%
  \BibitemOpen
  \bibfield  {author} {\bibinfo {author} {\bibfnamefont {S.}~\bibnamefont
  {Nadj-Perge}}, \bibinfo {author} {\bibfnamefont {I.~K.}\ \bibnamefont
  {Drozdov}}, \bibinfo {author} {\bibfnamefont {J.}~\bibnamefont {Li}},
  \bibinfo {author} {\bibfnamefont {H.}~\bibnamefont {Chen}}, \bibinfo {author}
  {\bibfnamefont {S.}~\bibnamefont {Jeon}}, \bibinfo {author} {\bibfnamefont
  {J.}~\bibnamefont {Seo}}, \bibinfo {author} {\bibfnamefont {A.~H.}\
  \bibnamefont {MacDonald}}, \bibinfo {author} {\bibfnamefont {B.~A.}\
  \bibnamefont {Bernevig}}, \ and\ \bibinfo {author} {\bibfnamefont
  {A.}~\bibnamefont {Yazdani}},\ }\href {\doibase 10.1126/science.1259327}
  {\bibfield  {journal} {\bibinfo  {journal} {Science}\ }\textbf {\bibinfo
  {volume} {346}},\ \bibinfo {pages} {602} (\bibinfo {year}
  {2014})}\BibitemShut {NoStop}%
\bibitem [{\citenamefont {Ruby}\ \emph {et~al.}(2015)\citenamefont {Ruby},
  \citenamefont {Pientka}, \citenamefont {Peng}, \citenamefont {von Oppen},
  \citenamefont {Heinrich},\ and\ \citenamefont {Franke}}]{Ruby2015}%
  \BibitemOpen
  \bibfield  {author} {\bibinfo {author} {\bibfnamefont {M.}~\bibnamefont
  {Ruby}}, \bibinfo {author} {\bibfnamefont {F.}~\bibnamefont {Pientka}},
  \bibinfo {author} {\bibfnamefont {Y.}~\bibnamefont {Peng}}, \bibinfo {author}
  {\bibfnamefont {F.}~\bibnamefont {von Oppen}}, \bibinfo {author}
  {\bibfnamefont {B.~W.}\ \bibnamefont {Heinrich}}, \ and\ \bibinfo {author}
  {\bibfnamefont {K.~J.}\ \bibnamefont {Franke}},\ }\href {\doibase
  10.1103/PhysRevLett.115.197204} {\bibfield  {journal} {\bibinfo  {journal}
  {Phys. Rev. Lett.}\ }\textbf {\bibinfo {volume} {115}},\ \bibinfo {pages}
  {197204} (\bibinfo {year} {2015})}\BibitemShut {NoStop}%
\bibitem [{\citenamefont {Kitaev}(2001)}]{Kitaev2001}%
  \BibitemOpen
  \bibfield  {author} {\bibinfo {author} {\bibfnamefont {A.~Y.}\ \bibnamefont
  {Kitaev}},\ }\href {http://stacks.iop.org/1063-7869/44/i=10S/a=S29}
  {\bibfield  {journal} {\bibinfo  {journal} {Phys. Usp.}\ }\textbf {\bibinfo
  {volume} {44}},\ \bibinfo {pages} {131} (\bibinfo {year} {2001})}\BibitemShut
  {NoStop}%
\bibitem [{\citenamefont {Mourik}\ \emph {et~al.}(2012)\citenamefont {Mourik},
  \citenamefont {Zuo}, \citenamefont {Frolov}, \citenamefont {Plissard},
  \citenamefont {Bakkers},\ and\ \citenamefont {Kouwenhoven}}]{Mourik2012}%
  \BibitemOpen
  \bibfield  {author} {\bibinfo {author} {\bibfnamefont {V.}~\bibnamefont
  {Mourik}}, \bibinfo {author} {\bibfnamefont {K.}~\bibnamefont {Zuo}},
  \bibinfo {author} {\bibfnamefont {S.~M.}\ \bibnamefont {Frolov}}, \bibinfo
  {author} {\bibfnamefont {S.~R.}\ \bibnamefont {Plissard}}, \bibinfo {author}
  {\bibfnamefont {E.~P. A.~M.}\ \bibnamefont {Bakkers}}, \ and\ \bibinfo
  {author} {\bibfnamefont {L.~P.}\ \bibnamefont {Kouwenhoven}},\ }\href
  {\doibase 10.1126/science.1222360} {\bibfield  {journal} {\bibinfo  {journal}
  {Science}\ }\textbf {\bibinfo {volume} {336}},\ \bibinfo {pages} {1003}
  (\bibinfo {year} {2012})}\BibitemShut {NoStop}%
\bibitem [{\citenamefont {Deng}\ \emph {et~al.}(2012)\citenamefont {Deng},
  \citenamefont {Yu}, \citenamefont {Huang}, \citenamefont {Larsson},
  \citenamefont {Caroff},\ and\ \citenamefont {Xu}}]{Deng2012}%
  \BibitemOpen
  \bibfield  {author} {\bibinfo {author} {\bibfnamefont {M.~T.}\ \bibnamefont
  {Deng}}, \bibinfo {author} {\bibfnamefont {C.~L.}\ \bibnamefont {Yu}},
  \bibinfo {author} {\bibfnamefont {G.~Y.}\ \bibnamefont {Huang}}, \bibinfo
  {author} {\bibfnamefont {M.}~\bibnamefont {Larsson}}, \bibinfo {author}
  {\bibfnamefont {P.}~\bibnamefont {Caroff}}, \ and\ \bibinfo {author}
  {\bibfnamefont {H.~Q.}\ \bibnamefont {Xu}},\ }\href {\doibase
  10.1021/nl303758w} {\bibfield  {journal} {\bibinfo  {journal} {Nano Lett.}\
  }\textbf {\bibinfo {volume} {12}},\ \bibinfo {pages} {6414} (\bibinfo {year}
  {2012})}\BibitemShut {NoStop}%
\bibitem [{\citenamefont {Das}\ \emph {et~al.}(2012)\citenamefont {Das},
  \citenamefont {Ronen}, \citenamefont {Most}, \citenamefont {Oreg},
  \citenamefont {Heiblum},\ and\ \citenamefont {Shtrikman}}]{das2012}%
  \BibitemOpen
  \bibfield  {author} {\bibinfo {author} {\bibfnamefont {A.}~\bibnamefont
  {Das}}, \bibinfo {author} {\bibfnamefont {Y.}~\bibnamefont {Ronen}}, \bibinfo
  {author} {\bibfnamefont {Y.}~\bibnamefont {Most}}, \bibinfo {author}
  {\bibfnamefont {Y.}~\bibnamefont {Oreg}}, \bibinfo {author} {\bibfnamefont
  {M.}~\bibnamefont {Heiblum}}, \ and\ \bibinfo {author} {\bibfnamefont
  {H.}~\bibnamefont {Shtrikman}},\ }\href
  {http://www.nature.com/nphys/journal/v8/n12/full/nphys2479.html} {\bibfield
  {journal} {\bibinfo  {journal} {Nat. Phys.}\ }\textbf {\bibinfo {volume}
  {8}},\ \bibinfo {pages} {887} (\bibinfo {year} {2012})}\BibitemShut {NoStop}%
\bibitem [{\citenamefont {Churchill}\ \emph {et~al.}(2013)\citenamefont
  {Churchill}, \citenamefont {Fatemi}, \citenamefont {Grove-Rasmussen},
  \citenamefont {Deng}, \citenamefont {Caroff}, \citenamefont {Xu},\ and\
  \citenamefont {Marcus}}]{Churchill2013}%
  \BibitemOpen
  \bibfield  {author} {\bibinfo {author} {\bibfnamefont {H.~O.~H.}\
  \bibnamefont {Churchill}}, \bibinfo {author} {\bibfnamefont {V.}~\bibnamefont
  {Fatemi}}, \bibinfo {author} {\bibfnamefont {K.}~\bibnamefont
  {Grove-Rasmussen}}, \bibinfo {author} {\bibfnamefont {M.~T.}\ \bibnamefont
  {Deng}}, \bibinfo {author} {\bibfnamefont {P.}~\bibnamefont {Caroff}},
  \bibinfo {author} {\bibfnamefont {H.~Q.}\ \bibnamefont {Xu}}, \ and\ \bibinfo
  {author} {\bibfnamefont {C.~M.}\ \bibnamefont {Marcus}},\ }\href {\doibase
  10.1103/PhysRevB.87.241401} {\bibfield  {journal} {\bibinfo  {journal} {Phys.
  Rev. B}\ }\textbf {\bibinfo {volume} {87}},\ \bibinfo {pages} {241401}
  (\bibinfo {year} {2013})}\BibitemShut {NoStop}%
\bibitem [{\citenamefont {Rokhinson}\ \emph {et~al.}(2012)\citenamefont
  {Rokhinson}, \citenamefont {Liu},\ and\ \citenamefont
  {Furdyna}}]{rokhinson2012}%
  \BibitemOpen
  \bibfield  {author} {\bibinfo {author} {\bibfnamefont {L.~P.}\ \bibnamefont
  {Rokhinson}}, \bibinfo {author} {\bibfnamefont {X.}~\bibnamefont {Liu}}, \
  and\ \bibinfo {author} {\bibfnamefont {J.~K.}\ \bibnamefont {Furdyna}},\
  }\href@noop {} {\bibfield  {journal} {\bibinfo  {journal} {Nat. Phys.}\
  }\textbf {\bibinfo {volume} {8}},\ \bibinfo {pages} {795} (\bibinfo {year}
  {2012})}\BibitemShut {NoStop}%
\bibitem [{\citenamefont {Albrecht}\ \emph {et~al.}(2016)\citenamefont
  {Albrecht}, \citenamefont {Higginbotham}, \citenamefont {Madsen},
  \citenamefont {Kuemmeth}, \citenamefont {Jespersen}, \citenamefont
  {Nyg{\aa}rd}, \citenamefont {Krogstrup},\ and\ \citenamefont
  {Marcus}}]{albrecht2016exponential}%
  \BibitemOpen
  \bibfield  {author} {\bibinfo {author} {\bibfnamefont {S.}~\bibnamefont
  {Albrecht}}, \bibinfo {author} {\bibfnamefont {A.}~\bibnamefont
  {Higginbotham}}, \bibinfo {author} {\bibfnamefont {M.}~\bibnamefont
  {Madsen}}, \bibinfo {author} {\bibfnamefont {F.}~\bibnamefont {Kuemmeth}},
  \bibinfo {author} {\bibfnamefont {T.}~\bibnamefont {Jespersen}}, \bibinfo
  {author} {\bibfnamefont {J.}~\bibnamefont {Nyg{\aa}rd}}, \bibinfo {author}
  {\bibfnamefont {P.}~\bibnamefont {Krogstrup}}, \ and\ \bibinfo {author}
  {\bibfnamefont {C.}~\bibnamefont {Marcus}},\ }\href
  {http://www.nature.com/nature/journal/v531/n7593/full/nature17162.html?WT.feed_name=subjects_electronics-photonics-and-device-physics}
  {\bibfield  {journal} {\bibinfo  {journal} {Nature}\ }\textbf {\bibinfo
  {volume} {531}},\ \bibinfo {pages} {206} (\bibinfo {year}
  {2016})}\BibitemShut {NoStop}%
\bibitem [{\citenamefont {Alicea}(2012)}]{alicea2012}%
  \BibitemOpen
  \bibfield  {author} {\bibinfo {author} {\bibfnamefont {J.}~\bibnamefont
  {Alicea}},\ }\href {http://stacks.iop.org/0034-4885/75/i=7/a=076501}
  {\bibfield  {journal} {\bibinfo  {journal} {Rep. Prog. Phys.}\ }\textbf
  {\bibinfo {volume} {75}},\ \bibinfo {pages} {076501} (\bibinfo {year}
  {2012})}\BibitemShut {NoStop}%
\bibitem [{\citenamefont {Li}\ and\ \citenamefont {Bai}(2013)}]{YuXianLi2013}%
  \BibitemOpen
  \bibfield  {author} {\bibinfo {author} {\bibfnamefont {Y.-X.}\ \bibnamefont
  {Li}}\ and\ \bibinfo {author} {\bibfnamefont {Z.-M.}\ \bibnamefont {Bai}},\
  }\href {\doibase 10.1063/1.4813229} {\bibfield  {journal} {\bibinfo
  {journal} {J. Appl. Phys.}\ }\textbf {\bibinfo {volume} {114}},\ \bibinfo
  {pages} {033703} (\bibinfo {year} {2013})}\BibitemShut {NoStop}%
\bibitem [{\citenamefont {Silva}\ and\ \citenamefont
  {Vernek}(2016)}]{Silva2016}%
  \BibitemOpen
  \bibfield  {author} {\bibinfo {author} {\bibfnamefont {J.~F.}\ \bibnamefont
  {Silva}}\ and\ \bibinfo {author} {\bibfnamefont {E.}~\bibnamefont {Vernek}},\
  }\href {http://stacks.iop.org/0953-8984/28/i=43/a=435702} {\bibfield
  {journal} {\bibinfo  {journal} {J. Phys. Condens. Matter}\ }\textbf {\bibinfo
  {volume} {28}},\ \bibinfo {pages} {435702} (\bibinfo {year}
  {2016})}\BibitemShut {NoStop}%
\bibitem [{\citenamefont {Barański}\ \emph {et~al.}(2017)\citenamefont
  {Barański}, \citenamefont {Kobiałka},\ and\ \citenamefont
  {Domański}}]{Baranski2017}%
  \BibitemOpen
  \bibfield  {author} {\bibinfo {author} {\bibfnamefont {J.}~\bibnamefont
  {Barański}}, \bibinfo {author} {\bibfnamefont {A.}~\bibnamefont
  {Kobiałka}}, \ and\ \bibinfo {author} {\bibfnamefont {T.}~\bibnamefont
  {Domański}},\ }\href {http://stacks.iop.org/0953-8984/29/i=7/a=075603}
  {\bibfield  {journal} {\bibinfo  {journal} {J. Phys. Condens. Matter}\
  }\textbf {\bibinfo {volume} {29}},\ \bibinfo {pages} {075603} (\bibinfo
  {year} {2017})}\BibitemShut {NoStop}%
\bibitem [{\citenamefont {Liu}\ and\ \citenamefont {Baranger}(2011)}]{Liu2011}%
  \BibitemOpen
  \bibfield  {author} {\bibinfo {author} {\bibfnamefont {D.~E.}\ \bibnamefont
  {Liu}}\ and\ \bibinfo {author} {\bibfnamefont {H.~U.}\ \bibnamefont
  {Baranger}},\ }\href {\doibase 10.1103/PhysRevB.84.201308} {\bibfield
  {journal} {\bibinfo  {journal} {Phys. Rev. B}\ }\textbf {\bibinfo {volume}
  {84}},\ \bibinfo {pages} {201308(R)} (\bibinfo {year} {2011})}\BibitemShut
  {NoStop}%
\bibitem [{\citenamefont {Vernek}\ \emph {et~al.}(2014)\citenamefont {Vernek},
  \citenamefont {Penteado}, \citenamefont {Seridonio},\ and\ \citenamefont
  {Egues}}]{Vernek2014}%
  \BibitemOpen
  \bibfield  {author} {\bibinfo {author} {\bibfnamefont {E.}~\bibnamefont
  {Vernek}}, \bibinfo {author} {\bibfnamefont {P.~H.}\ \bibnamefont
  {Penteado}}, \bibinfo {author} {\bibfnamefont {A.~C.}\ \bibnamefont
  {Seridonio}}, \ and\ \bibinfo {author} {\bibfnamefont {J.~C.}\ \bibnamefont
  {Egues}},\ }\href {\doibase 10.1103/PhysRevB.89.165314} {\bibfield  {journal}
  {\bibinfo  {journal} {Phys. Rev. B}\ }\textbf {\bibinfo {volume} {89}},\
  \bibinfo {pages} {165314} (\bibinfo {year} {2014})}\BibitemShut {NoStop}%
\bibitem [{\citenamefont {Deng}\ \emph {et~al.}(2016)\citenamefont {Deng},
  \citenamefont {Vaitiekenas}, \citenamefont {Hansen}, \citenamefont {Danon},
  \citenamefont {Leijnse}, \citenamefont {Flensberg}, \citenamefont {Nyg{\r
  a}rd}, \citenamefont {Krogstrup},\ and\ \citenamefont {Marcus}}]{Deng2016}%
  \BibitemOpen
  \bibfield  {author} {\bibinfo {author} {\bibfnamefont {M.~T.}\ \bibnamefont
  {Deng}}, \bibinfo {author} {\bibfnamefont {S.}~\bibnamefont {Vaitiekenas}},
  \bibinfo {author} {\bibfnamefont {E.~B.}\ \bibnamefont {Hansen}}, \bibinfo
  {author} {\bibfnamefont {J.}~\bibnamefont {Danon}}, \bibinfo {author}
  {\bibfnamefont {M.}~\bibnamefont {Leijnse}}, \bibinfo {author} {\bibfnamefont
  {K.}~\bibnamefont {Flensberg}}, \bibinfo {author} {\bibfnamefont
  {J.}~\bibnamefont {Nyg{\r a}rd}}, \bibinfo {author} {\bibfnamefont
  {P.}~\bibnamefont {Krogstrup}}, \ and\ \bibinfo {author} {\bibfnamefont
  {C.~M.}\ \bibnamefont {Marcus}},\ }\href {\doibase 10.1126/science.aaf3961}
  {\bibfield  {journal} {\bibinfo  {journal} {Science}\ }\textbf {\bibinfo
  {volume} {354}},\ \bibinfo {pages} {1557} (\bibinfo {year}
  {2016})}\BibitemShut {NoStop}%
\bibitem [{\citenamefont {Ueda}\ and\ \citenamefont
  {Yokoyama}(2014)}]{Ueda2014}%
  \BibitemOpen
  \bibfield  {author} {\bibinfo {author} {\bibfnamefont {A.}~\bibnamefont
  {Ueda}}\ and\ \bibinfo {author} {\bibfnamefont {T.}~\bibnamefont
  {Yokoyama}},\ }\href {\doibase 10.1103/PhysRevB.90.081405} {\bibfield
  {journal} {\bibinfo  {journal} {Phys. Rev. B}\ }\textbf {\bibinfo {volume}
  {90}},\ \bibinfo {pages} {081405} (\bibinfo {year} {2014})}\BibitemShut
  {NoStop}%
\bibitem [{\citenamefont {Xia}\ \emph {et~al.}(2015)\citenamefont {Xia},
  \citenamefont {Duan},\ and\ \citenamefont {Zhang}}]{Xia2015}%
  \BibitemOpen
  \bibfield  {author} {\bibinfo {author} {\bibfnamefont {J.-J.}\ \bibnamefont
  {Xia}}, \bibinfo {author} {\bibfnamefont {S.-Q.}\ \bibnamefont {Duan}}, \
  and\ \bibinfo {author} {\bibfnamefont {W.}~\bibnamefont {Zhang}},\ }\href
  {\doibase 10.1186/s11671-015-0914-3} {\bibfield  {journal} {\bibinfo
  {journal} {‎Nanoscale Res. Lett.}\ }\textbf {\bibinfo {volume} {10}},\
  \bibinfo {pages} {223} (\bibinfo {year} {2015})}\BibitemShut {NoStop}%
\bibitem [{\citenamefont {En-Ming}\ \emph {et~al.}(2014)\citenamefont
  {En-Ming}, \citenamefont {Yi-Ming}, \citenamefont {Lu-Bing},\ and\
  \citenamefont {Bai-Gen}}]{EnMing2014}%
  \BibitemOpen
  \bibfield  {author} {\bibinfo {author} {\bibfnamefont {S.}~\bibnamefont
  {En-Ming}}, \bibinfo {author} {\bibfnamefont {P.}~\bibnamefont {Yi-Ming}},
  \bibinfo {author} {\bibfnamefont {S.}~\bibnamefont {Lu-Bing}}, \ and\
  \bibinfo {author} {\bibfnamefont {W.}~\bibnamefont {Bai-Gen}},\ }\href
  {http://stacks.iop.org/1674-1056/23/i=5/a=057201} {\bibfield  {journal}
  {\bibinfo  {journal} {Chin. Phys. B}\ }\textbf {\bibinfo {volume} {23}},\
  \bibinfo {pages} {057201} (\bibinfo {year} {2014})}\BibitemShut {NoStop}%
\bibitem [{\citenamefont {Ricco}\ \emph
  {et~al.}(2016{\natexlab{a}})\citenamefont {Ricco}, \citenamefont {Marques},
  \citenamefont {Dessotti}, \citenamefont {de~Souza},\ and\ \citenamefont
  {Seridonio}}]{Ricco20162}%
  \BibitemOpen
  \bibfield  {author} {\bibinfo {author} {\bibfnamefont {L.}~\bibnamefont
  {Ricco}}, \bibinfo {author} {\bibfnamefont {Y.}~\bibnamefont {Marques}},
  \bibinfo {author} {\bibfnamefont {F.}~\bibnamefont {Dessotti}}, \bibinfo
  {author} {\bibfnamefont {M.}~\bibnamefont {de~Souza}}, \ and\ \bibinfo
  {author} {\bibfnamefont {A.}~\bibnamefont {Seridonio}},\ }\href {\doibase
  10.1016/j.physe.2015.11.025} {\bibfield  {journal} {\bibinfo  {journal}
  {Phys. E}\ }\textbf {\bibinfo {volume} {78}},\ \bibinfo {pages} {25 }
  (\bibinfo {year} {2016}{\natexlab{a}})}\BibitemShut {NoStop}%
\bibitem [{\citenamefont {Zeng}\ \emph {et~al.}(2016)\citenamefont {Zeng},
  \citenamefont {Chen},\ and\ \citenamefont {Lü}}]{Zeng2016}%
  \BibitemOpen
  \bibfield  {author} {\bibinfo {author} {\bibfnamefont {Q.-B.}\ \bibnamefont
  {Zeng}}, \bibinfo {author} {\bibfnamefont {S.}~\bibnamefont {Chen}}, \ and\
  \bibinfo {author} {\bibfnamefont {R.}~\bibnamefont {Lü}},\ }\href {\doibase
  10.1016/j.physleta.2015.12.026} {\bibfield  {journal} {\bibinfo  {journal}
  {Phys. Lett. A}\ }\textbf {\bibinfo {volume} {380}},\ \bibinfo {pages} {951 }
  (\bibinfo {year} {2016})}\BibitemShut {NoStop}%
\bibitem [{\citenamefont {Ricco}\ \emph
  {et~al.}(2016{\natexlab{b}})\citenamefont {Ricco}, \citenamefont {Marques},
  \citenamefont {Dessotti}, \citenamefont {Machado}, \citenamefont {de~Souza},\
  and\ \citenamefont {Seridonio}}]{PhysRevB.93.165116}%
  \BibitemOpen
  \bibfield  {author} {\bibinfo {author} {\bibfnamefont {L.~S.}\ \bibnamefont
  {Ricco}}, \bibinfo {author} {\bibfnamefont {Y.}~\bibnamefont {Marques}},
  \bibinfo {author} {\bibfnamefont {F.~A.}\ \bibnamefont {Dessotti}}, \bibinfo
  {author} {\bibfnamefont {R.~S.}\ \bibnamefont {Machado}}, \bibinfo {author}
  {\bibfnamefont {M.}~\bibnamefont {de~Souza}}, \ and\ \bibinfo {author}
  {\bibfnamefont {A.~C.}\ \bibnamefont {Seridonio}},\ }\href {\doibase
  10.1103/PhysRevB.93.165116} {\bibfield  {journal} {\bibinfo  {journal} {Phys.
  Rev. B}\ }\textbf {\bibinfo {volume} {93}},\ \bibinfo {pages} {165116}
  (\bibinfo {year} {2016}{\natexlab{b}})}\BibitemShut {NoStop}%
\bibitem [{\citenamefont {Flensberg}(2011)}]{Flensberg2011}%
  \BibitemOpen
  \bibfield  {author} {\bibinfo {author} {\bibfnamefont {K.}~\bibnamefont
  {Flensberg}},\ }\href {\doibase 10.1103/PhysRevLett.106.090503} {\bibfield
  {journal} {\bibinfo  {journal} {Phys. Rev. Lett.}\ }\textbf {\bibinfo
  {volume} {106}},\ \bibinfo {pages} {090503} (\bibinfo {year}
  {2011})}\BibitemShut {NoStop}%
\bibitem [{\citenamefont {Hyart}\ \emph {et~al.}(2013)\citenamefont {Hyart},
  \citenamefont {van Heck}, \citenamefont {Fulga}, \citenamefont {Burrello},
  \citenamefont {Akhmerov},\ and\ \citenamefont {Beenakker}}]{Hyart2013}%
  \BibitemOpen
  \bibfield  {author} {\bibinfo {author} {\bibfnamefont {T.}~\bibnamefont
  {Hyart}}, \bibinfo {author} {\bibfnamefont {B.}~\bibnamefont {van Heck}},
  \bibinfo {author} {\bibfnamefont {I.~C.}\ \bibnamefont {Fulga}}, \bibinfo
  {author} {\bibfnamefont {M.}~\bibnamefont {Burrello}}, \bibinfo {author}
  {\bibfnamefont {A.~R.}\ \bibnamefont {Akhmerov}}, \ and\ \bibinfo {author}
  {\bibfnamefont {C.~W.~J.}\ \bibnamefont {Beenakker}},\ }\href {\doibase
  10.1103/PhysRevB.88.035121} {\bibfield  {journal} {\bibinfo  {journal} {Phys.
  Rev. B}\ }\textbf {\bibinfo {volume} {88}},\ \bibinfo {pages} {035121}
  (\bibinfo {year} {2013})}\BibitemShut {NoStop}%
\bibitem [{\citenamefont {Li}\ \emph {et~al.}(2016)\citenamefont {Li},
  \citenamefont {Neupert}, \citenamefont {Bernevig},\ and\ \citenamefont
  {Yazdani}}]{li2016}%
  \BibitemOpen
  \bibfield  {author} {\bibinfo {author} {\bibfnamefont {J.}~\bibnamefont
  {Li}}, \bibinfo {author} {\bibfnamefont {T.}~\bibnamefont {Neupert}},
  \bibinfo {author} {\bibfnamefont {B.~A.}\ \bibnamefont {Bernevig}}, \ and\
  \bibinfo {author} {\bibfnamefont {A.}~\bibnamefont {Yazdani}},\ }\href
  {http://www.nature.com/articles/ncomms10395} {\bibfield  {journal} {\bibinfo
  {journal} {Nat. commun.}\ }\textbf {\bibinfo {volume} {7}} (\bibinfo {year}
  {2016})}\BibitemShut {NoStop}%
\bibitem [{\citenamefont {Aristov}\ and\ \citenamefont
  {Gutman}(2016)}]{Aristov2016}%
  \BibitemOpen
  \bibfield  {author} {\bibinfo {author} {\bibfnamefont {D.~N.}\ \bibnamefont
  {Aristov}}\ and\ \bibinfo {author} {\bibfnamefont {D.~B.}\ \bibnamefont
  {Gutman}},\ }\href {http://stacks.iop.org/1751-8121/49/i=31/a=315301}
  {\bibfield  {journal} {\bibinfo  {journal} {J. Phys. A Math Theor.}\ }\textbf
  {\bibinfo {volume} {49}},\ \bibinfo {pages} {315301} (\bibinfo {year}
  {2016})}\BibitemShut {NoStop}%
\bibitem [{\citenamefont {Flensberg}(2010)}]{Flensberg2010}%
  \BibitemOpen
  \bibfield  {author} {\bibinfo {author} {\bibfnamefont {K.}~\bibnamefont
  {Flensberg}},\ }\href {\doibase 10.1103/PhysRevB.82.180516} {\bibfield
  {journal} {\bibinfo  {journal} {Phys. Rev. B}\ }\textbf {\bibinfo {volume}
  {82}},\ \bibinfo {pages} {180516} (\bibinfo {year} {2010})}\BibitemShut
  {NoStop}%
\bibitem [{\citenamefont {Ruiz-Tijerina}\ \emph {et~al.}(2015)\citenamefont
  {Ruiz-Tijerina}, \citenamefont {Vernek}, \citenamefont {Dias~da Silva},\ and\
  \citenamefont {Egues}}]{ruiz2015}%
  \BibitemOpen
  \bibfield  {author} {\bibinfo {author} {\bibfnamefont {D.~A.}\ \bibnamefont
  {Ruiz-Tijerina}}, \bibinfo {author} {\bibfnamefont {E.}~\bibnamefont
  {Vernek}}, \bibinfo {author} {\bibfnamefont {L.~G. G.~V.}\ \bibnamefont
  {Dias~da Silva}}, \ and\ \bibinfo {author} {\bibfnamefont {J.~C.}\
  \bibnamefont {Egues}},\ }\href {\doibase 10.1103/PhysRevB.91.115435}
  {\bibfield  {journal} {\bibinfo  {journal} {Phys. Rev. B}\ }\textbf {\bibinfo
  {volume} {91}},\ \bibinfo {pages} {115435} (\bibinfo {year}
  {2015})}\BibitemShut {NoStop}%
\bibitem [{\citenamefont {Bolech}\ and\ \citenamefont
  {Demler}(2007)}]{Bolech2007}%
  \BibitemOpen
  \bibfield  {author} {\bibinfo {author} {\bibfnamefont {C.~J.}\ \bibnamefont
  {Bolech}}\ and\ \bibinfo {author} {\bibfnamefont {E.}~\bibnamefont
  {Demler}},\ }\href {\doibase 10.1103/PhysRevLett.98.237002} {\bibfield
  {journal} {\bibinfo  {journal} {Phys. Rev. Lett.}\ }\textbf {\bibinfo
  {volume} {98}},\ \bibinfo {pages} {237002} (\bibinfo {year}
  {2007})}\BibitemShut {NoStop}%
\bibitem [{\citenamefont {Leijnse}\ and\ \citenamefont
  {Flensberg}(2011)}]{Leijnse2011}%
  \BibitemOpen
  \bibfield  {author} {\bibinfo {author} {\bibfnamefont {M.}~\bibnamefont
  {Leijnse}}\ and\ \bibinfo {author} {\bibfnamefont {K.}~\bibnamefont
  {Flensberg}},\ }\href {\doibase 10.1103/PhysRevLett.107.210502} {\bibfield
  {journal} {\bibinfo  {journal} {Phys. Rev. Lett.}\ }\textbf {\bibinfo
  {volume} {107}},\ \bibinfo {pages} {210502} (\bibinfo {year}
  {2011})}\BibitemShut {NoStop}%
\bibitem [{\citenamefont {L\"u}\ \emph {et~al.}(2012)\citenamefont {L\"u},
  \citenamefont {Lu},\ and\ \citenamefont {Shen}}]{Lu2012}%
  \BibitemOpen
  \bibfield  {author} {\bibinfo {author} {\bibfnamefont {H.-F.}\ \bibnamefont
  {L\"u}}, \bibinfo {author} {\bibfnamefont {H.-Z.}\ \bibnamefont {Lu}}, \ and\
  \bibinfo {author} {\bibfnamefont {S.-Q.}\ \bibnamefont {Shen}},\ }\href
  {\doibase 10.1103/PhysRevB.86.075318} {\bibfield  {journal} {\bibinfo
  {journal} {Phys. Rev. B}\ }\textbf {\bibinfo {volume} {86}},\ \bibinfo
  {pages} {075318} (\bibinfo {year} {2012})}\BibitemShut {NoStop}%
\end{thebibliography}%

\end{document}